\documentstyle[aps,floats,epsfig]{revtex}
\newcommand{\ead}[1]{\vspace*{5pt}\address{E-mail: #1}}
\newcommand{\ack}{\section*{Acknowledgements}}

\newcommand{\p}{\alpha}
\newcommand{\q}{\beta}
\newcommand{\m}{\chi}
\newcommand{\n}{\sigma}
\newcommand{\x}{\kappa}
\newcommand{\y}{\nu}
\newcommand{\Ho}{\hat{H}_0}
\newcommand{\Hsp}{\hat{H}_{sp}}
\newcommand{\Hi}{\hat{H_I}}

\begin{document}

\draft
\title{Dynamics of thermal Bose fields in the classical limit}
\author{M J Davis\dag, R J Ballagh\ddag\  and K Burnett\dag}
\address{\dag\ Clarendon Laboratory, Department of Physics, University of Oxford,
Oxford OX1~3PU, UK\\
\ddag\ Department of Physics, University of Otago, PO Box 56, Dunedin, New
Zealand}

\maketitle

\ead{m.davis2@physics.ox.ac.uk}

\begin{abstract}

We develop an approximate formalism suitable for performing simulations of the
thermal dynamics of interacting Bose gases.  The method is based on the
observation that when the lowest energy modes of the Bose field operator are
highly occupied, they may be treated classically to a good approximation.  We
derive a finite temperature Gross-Pitaevskii equation for these modes which is 
coupled to an effective reservoir described by quantum kinetic theory.  We
discuss each of the terms that arise in this Gross-Pitaevskii equation, and their
relevance to experimental systems.  We then describe a simpler projected
Gross-Pitaevskii equation that may be useful in simulating thermal Bose
condensates. This classical method could be applied to other Bose fields.

\end{abstract}


\section{Introduction}
 
The achievement of Bose-Einstein condensation (BEC) in a dilute gas 
\cite{JILA,MIT,RICE} offers the possibility of studying the dynamics of a
quantum field in the laboratory.  In principle it presents an opportunity to
directly compare computational predictions with experimental results;  however
carrying out dynamical calculations of thermal quantum fields is an extremely
difficult problem that generally requires severe approximations to be made.

The most successful finite temperature theories of BEC are based on second-order
perturbation theory, and are limited to the case of thermal equilibrium away
from the region of critical fluctuations  \cite{sam,fedichev,giorgini}.  These
theories  have allowed the accurate determination from first principles of 
 quantities such as excitation frequencies and
damping rates of Bose-condensed systems.  However, their
extension to dynamical situations is computationally difficult.

At very low temperatures when most of the atoms are in the condensate, the
Gross-Pitaevskii equation (GPE) has proved remarkably successful in numerically
modelling BEC experiments.  
The time-dependent GPE has the form
\begin{equation}
i \hbar \frac{\partial {\psi}({\bf x})}{\partial t}
=\Hsp \psi({\bf x}) + U_0 |\psi({\bf x})|^2 \psi({\bf x}),
\label{eqn:gpe}
\end{equation}
where  $U_0 = 4\pi \hbar^2 a / m$ is
the effective interaction strength at low momenta,  $a$ is the {\em s}-wave
scattering length, and $m$ is the particle mass. $\Hsp$ is the 
single particle Hamiltonian
\begin{equation}
\Hsp = -\frac{\hbar^2}{2m}\nabla^2 + V_{\rm trap}({\bf x}),
\label{eqn:Hsp}
\end{equation}
where $V_{\rm trap}({\bf x})$ is the external confining 
potential of the system. 
The GPE can be derived by a number of different approaches (e.g.\ a 
number-conserving approach \cite{truebog}), but a direct method is to take the mean 
mean value of the equation of motion for the Bose field operator and assume that
the quantum fluctuations can be neglected. This procedure effectively assumes
that the field is well approximated by a coherent state.  

It has been argued that the GPE can also describe the dynamics of a Bose-Einstein
condensate at finite temperature \cite{boris,kagan1,kagan2,kagan3} for the reason
that in the limit that all the low-lying modes of the system are highly occupied
($N_k\gg1$), the classical fluctuations of the field operator $\hat{\Psi}({\bf
x},t)$  will be much larger than the quantum fluctuations. It is therefore
reasonable to neglect the quantum fluctuations, and thus all highly-occupied modes
may be well approximated by a coherent wave function.  This is analogous to the
situation in laser physics where the highly occupied laser modes can be well
described by classical equations.   Despite this proposal appearing in the
literature in 1991 \cite{boris}, very few numerical calculations have been
performed. The first was by Damle {\em et al.}\ \cite{damle}, who calculated the
approach to equilibrium of a near-ideal superfluid using the GPE.  Subsequently,
Marshall {\em et al.}\ \cite{Marshall} performed two-dimensional simulations of
evaporative cooling of a thermal Bose field  in a trap.  More recently, papers by
Stoof and Bijlsma \cite{stoof_dynamics} and Sinatra {\em et al.}\
\cite{sinatra_short} have used classical methods in dynamical calculations of
thermal one-dimensional Bose-Einstein condensates.   Goral {\em et al.}\
\cite{Goral} have performed dynamical calculations that treat several modes of a
3D homogeneous Bose gas classically.  Their method, while not specifically employing
the GPE, is similar to the approach we suggest here and for which we have
presented our own numerical results in reference \cite{ftgpe}. Classical
approximations to other quantum field equations have also been successful in the
calculation of the dynamics of the electroweak phase transition \cite{Turok}.  

The major advantage of using the GPE to describe thermal dynamics is simply
that, while it is still a major computational task,  it is possible to
numerically solve the equation for realistic systems in a reasonable amount of
time.  In addition, the GPE is non-perturbative and it should be possible to
study the region of the BEC phase transition, where perturbation theory often
fails.

There are, however, a number of problems associated with the use of the GPE to
represent the entire Bose field at finite temperature.  It is a classical
equation, and so in equilibrium it will satisfy the equipartition theorem---all
modes of the system will have an occupation of $N_k = k_B T/\epsilon_k$.  Thus,
if we couple the GPE to a heat bath and numerically solve the equation with
infinite accuracy, we will observe an ultra-violet catastrophe.  Also, we can
see that the higher the energy of any given mode, the lower its occupation will
be in equilibrium and at a sufficiently high energy the criterion  $N_k \gg 1$
will no longer be satisfied.  For these low occupation modes a form of kinetic
equation is more appropriate.  The solution to both of these problems is to
introduce a {\em cutoff} in the modes represented by the GPE.

In this paper we develop an approximate formalism in which the low-lying modes
of the system are described non-perturbatively by the GPE, coupled to a
thermal bath described by a quantum Boltzmann equation.  We derive a finite
temperature Gross-Pitaevskii equation (FTGPE), and discuss the terms that
couple the part of the field operator represented by a coherent wave function
to the thermal bath.  In particular we show how a description of loss via
elastic collisions arises naturally in the formalism. 

Several other authors have developed formalisms for non-equilibrium dynamics  using
quite different theoretical methods---we mention  Gardiner and
Zoller \cite{QKI,QKIII,QKV}, Proukakis {\em et al.}\ \cite{prou}, Stoof
\cite{stoof_prl,stoof_ft}, Zaremba {\em et al.}\ \cite{ZGN}, Walser {\em et al.}\
\cite{walser}, and finally Sinatra {\em et al.}\ \cite{sinatra}.  The work we
present here has elements in common with several of these.  In particular, the
average of the quantum Langevin equation written down in the formalism of Stoof \cite{stoof_prl,stoof_ft}
would correspond to the FTGPE we derive in section \ref{eom}. 

This paper is organized as follows.  In section~\ref{hamiltonian} we write down
and discuss the Hamiltonian which is our starting point. In
section~\ref{projection} we outline how we decompose the field operator into a
coherent and incoherent region, and in section~\ref{eom} we derive a finite
temperature Gross-Pitaevskii equation that forms the main result of this
paper.  In  section~\ref{ftgpe_terms} we discuss the terms that arise in the
FTGPE, their relation to experiments, and their approximate form in terms of
occupations numbers of incoherent region modes.  We discuss a simple finite
temperature equation which we call the projected GPE in section~\ref{pgpe} and
finally conclude in section~\ref{conclusions}.

\section{Hamiltonian}\label{hamiltonian}

We begin with the usual second quantized many-body Hamiltonian for a system of identical,
structureless bosons with pair-wise interactions
\begin{eqnarray}
&&\hat{H} = \Ho + \Hi,
\label{eqn:H}
\\
&&\Ho= \int d^3{\bf x}\; \hat{\Psi}^{\dag}({\bf x},t)
\Hsp
\hat{\Psi}({\bf x},t),
\label{eqn:Ho}
\\
&&\Hi= \frac{1}{2}\int d^3{\bf x} \int d^3{\bf x}' \;
\hat{\Psi}^{\dag}({\bf x},t) \hat{\Psi}^{\dag}({\bf x}',t)
V({\bf x}-{\bf x}')
\hat{\Psi}({\bf x}',t)\hat{\Psi}({\bf x},t).
\label{eqn:Hi}
\end{eqnarray}
The non-interacting part of the Hamiltonian, $\Ho$, corresponds to an ideal
gas and can 
be diagonalized exactly by the eigenvectors of $\Hsp$. 
The quantity $\Hi$ describes 
two-body interactions via the interatomic potential $V({\bf x})$.
The field operator $\hat{\Psi}({\bf x},t)$ annihilates a single boson  of mass $m$
at position ${\bf x}$ and time $t$, and obeys the equal-time commutation relations
\begin{equation}
\left[\hat{\Psi}({\bf x},t) , \hat{\Psi}({\bf x}',t)\right]
=
\left[\hat{\Psi}^{\dag}({\bf x},t) , \hat{\Psi}^{\dag}({\bf x}',t)\right]
= 0,
\end{equation}
\begin{equation}
\left[\hat{\Psi}({\bf x},t) , \hat{\Psi}^{\dag}({\bf x}',t)\right]
= \delta ({\bf x}-{\bf x}').
\label{eqn:Psicommute}
\end{equation}
The field operator is normalised such that 
\begin{equation}
\int d^3{\bf x} \;\hat{\Psi}^{\dag}({\bf x},t) \hat{\Psi}({\bf x},t)
=\hat{N},
\label{eqn:fieldnorm}
\end{equation}
where $\hat{N}$ is the particle number operator of the system.

\subsection{Basis set representation}

It is useful to expand the field operator on a basis set
\begin{equation}
\hat{\Psi}({\bf x},t) = \sum_{n} \hat{a}_{ n}(t) \phi_{ n}({\bf x}),
\label{eqn:field_expand}
\end{equation}
where $\phi_{ n}({\bf x})$ is a mode function, and 
$\hat{a}_{ n}(t)$ annihilates a particle in mode ${n}$ at time $t$. These
operators obey the equal-time commutation relations
\begin{equation}
\left[\hat{a}_m , \hat{a}_n\right]
=
\left[\hat{a}_m^{\dag} , \hat{a}_n^{\dag}\right]
= 0,
\end{equation}
\begin{equation}
\left[\hat{a}_m , \hat{a}_n^{\dag}\right]
= \delta_{mn}.
\label{eqn:commute}
\end{equation}
where we have dropped the time labels for clarity.
If we
substitute equation~(\ref{eqn:field_expand}) into the Hamiltonian (\ref{eqn:H}) and take the 
set $\{\phi_{ n}\}$ to be the eigenvectors of $\Hsp$, we find
\begin{eqnarray}
\hat{H}& = & 
\sum_{n} \hbar \omega_{n}
\hat{a}^{\dag}_{n}\hat{a}_{n}
+\frac{1}{2}
\sum_{pqmn}\langle pq|V|mn \rangle
\hat{a}^{\dag}_{p}\hat{a}^{\dag}_{q}\hat{a}_{m}\hat{a}_{n},
\label{eqn:Hbasis}
\end{eqnarray}
where we have defined the symmetrized matrix element
\begin{eqnarray}
\langle pq|V|mn \rangle
&=&
\frac{1}{2}\int d^3{\bf x} \int d^3{\bf x}'\;
\phi^*_{p}({\bf x}) \phi^*_{q}({\bf x}')
V({\bf x}-{\bf x}')
\phi_{m}({\bf x}') \phi_{n}({\bf x})
\nonumber \\
&+&
\frac{1}{2}\int d^3{\bf x} \int d^3{\bf x}'\;
\phi^*_{p}({\bf x}) \phi^*_{q}({\bf x}')
V({\bf x}-{\bf x}')
 \phi_{n}({\bf x}')\phi_{m}({\bf x}).
\label{eqn:matrix_el}
\end{eqnarray}
whose use  significantly reduces the length of the  equations of motion we later
derive.  Equation~(\ref{eqn:matrix_el}) represents both direct and exchange
collisions, which are physically indistinguishable for identical bosons.

The Heisenberg equation of motion for the individual mode operator
 $\hat{a}_p$ is therefore
\begin{equation}
i \hbar \frac{d \hat{a}_p}{dt}
=
\hbar \omega_p \hat{a}_p + 
\sum_{qmn}\langle pq|V|mn\rangle
\hat{a}_q^{\dag}\hat{a}_m\hat{a}_n,
\label{eqn:a}
\end{equation}
and we can define slowly-varying operators
\begin{equation}
\tilde{a}_p = \hat{a}_p e^{i \omega_p t},
\end{equation}
so that the equation of motion for the annihilation operator becomes
\begin{equation}
i \hbar \frac{d \tilde{a}_p}{dt}
=
\sum_{qmn}\langle pq|V|mn\rangle
\tilde{a}_q^{\dag}\tilde{a}_m\tilde{a}_n
e^{i(\omega_p + \omega_q - \omega_m - \omega_n)t}.
\label{eqn:at}
\end{equation}

\subsection{Effective low-energy Hamiltonian}\label{eff_lowH}

The Hamiltonian described above contains spatial integrals over the bare
interatomic potential  between two atoms, $V({\bf x})$.  However, it is
well-known that at low temperatures the scattering of neutral atoms in three
dimensions can be described by the {\em s}-wave scattering length $a$.  This
parameter is often introduced into the theory by replacing the real interatomic
potential by the contact potential
\begin{equation}
V({\bf x}-{\bf x}') \rightarrow U_0 \delta({\bf x}-{\bf x}'), \;\;\;\;\;
U_0 = \frac{4 \pi \hbar^2 a}{m}.
\label{eqn:U0_approx}
\end{equation}
The interaction strength $U_0$ can be shown to arise from
 the increase in kinetic energy of a
two-particle wave function, when an excluded region of radius $a$ is introduced
corresponding to a hard
sphere interaction potential \cite{keith_varenna}.
The contact potential approximation, however, can lead to 
ultraviolet divergences in theories
of BEC if it is simply substituted into the Hamiltonian (\ref{eqn:H}).
 This is not surprising, as the delta-function
potential can scatter high-energy atoms just as effectively as low-energy atoms.
Physically this is unrealistic, as momentum transfer between atoms will vanish
at high momenta $(k>1/a)$.  The contact potential is a low-energy approximation,
and care must be taken when summing over high energy states.

The ultraviolet renormalization of the theory can be achieved by introducing
the two-body T-matrix into the Hamiltonian, resulting in a high-momentum cutoff
$K$ to the the states considered \cite{sam}.  This procedure is valid as long
as the condition $Ka \ll 1$ is satisfied for the entire gas.  We will consider
the issue of ultraviolet divergence further in
section~\ref{sec:linear}.

\section{Projection operator}\label{projection} 

The aim of this paper is to represent the highly occupied modes of the field
operator $\hat{\Psi}({\bf x})$ by a wave function $\psi({\bf x})$.  It is
reasonable to neglect the quantum fluctuations of these modes, and therefore
$\psi({\bf x})$ represents a classical field. To achieve our goal we divide our
representation of the field operator into two separate regions.  The first, in
which the condition $N_k \gg 1$ is satisfied, we call the coherent region $C$. 
The other which contains the remainder of the field, is denoted the incoherent
region $I$.

We define the coherent region projection operator
\begin{eqnarray}
\hat{\mathcal{P}} & =& \sum_{\nu \in C} | \nu \rangle \langle \nu |,
\label{eqn:projector}
\end{eqnarray}
where the region $C$ is {\em determined} by the 
requirement that $\langle \hat{a}_{\nu}^{\dag} \hat{a}_{\nu}\rangle \gg 1$,
and the set $|\nu \rangle$ defines some basis in which the field operator is
approximately diagonal near the cutoff.  This condition is imposed simply
so that in equilibrium  the quantity
$\langle \hat{a}_{\nu}^{\dag} \hat{a}_{\nu}\rangle$ has a well-defined average
at the boundary of $C$ that can be interpreted as a mode occupation number.
The position of the cutoff is a choice that must be made before any
calculation, and is required for the construction of the initial wave
function. 

Operating on the field operator with $\hat{\mathcal{P}}$ gives
\begin{eqnarray}
\hat{\mathcal{P}}\hat{\Psi}({\bf x})  &= &
\sum_{\nu \in C} \phi_{\nu}({\bf x}) \int d^3 {\bf x}'\; \phi^*_{
 \nu}({\bf x}')\hat{\Psi}({\bf x}'),\nonumber\\
&=&
\sum_{{ \nu} \in C} \hat{a}_{ \nu} \phi_{ \nu}({\bf x}),\nonumber\\
&\equiv&
\hat{\psi}({\bf x}).
\end{eqnarray}
such that $\hat{\psi}({\bf x})$ is the field operator for the coherent region.
We now introduce the orthogonal projector 
$\hat{\mathcal{Q}} = \hat{1} - \hat{\mathcal{P}}$ and define
\begin{eqnarray} 
\hat{\mathcal{Q}}\hat{\Psi}({\bf x})
&=& \sum_{{ k} \notin C} \hat{a}_{ k} \phi_{ k}({\bf x}),\nonumber\\
 &\equiv&
  \hat{\eta}({\bf x}).
\label{eqn:eta}
\end{eqnarray}
The quantity $\hat{\eta}({\bf x})$ is the field operator for the incoherent
region and represents an effective
thermal bath.  Quantum fluctuations {\em are} important for these
modes---in fact we
will later assume that $\langle \hat{a}_{ k} \rangle \approx 0$ for the large
majority of ${k} \notin C$.

The full field operator is
\begin{eqnarray}
\hat{\Psi}({\bf x}) 
&=& 
[\hat{\mathcal{P}} + \hat{\mathcal{Q}}]\hat{\Psi}({\bf x}), \nonumber \\
 &=& 
\hat{\psi}({\bf x}) + \hat{\eta}({\bf x}),\nonumber\\
&=&
\sum_{{ \nu} \in C} \hat{a}_{ \nu} \phi_{ \nu}({\bf x})+ 
\sum_{{ k} \notin C} \hat{a}_{k} \phi_{ k}({\bf x}),
\label{eqn:fieldsplit}
\end{eqnarray}
where we indicate indices within $C$ by Greek subscripts, and outside $C$ by
Roman subscripts.  We shall follow this convention throughout the body of
this paper.

As an example of the size of the regions we consider a $^{87}$Rb gas in a
harmonic trap with a geometric mean trap frequency of $\bar{\omega} = 2\pi
\times 100$ Hz.  For a condensate of $10^6$ atoms at a temperature of $640$ nK
(and hence a total number of trapped atoms of about $5 \times 10^6$), we find
that a quantum level with energy  $\epsilon - \mu \approx 15 \hbar
\bar{\omega}$ has a mean occupation of  $\langle N \rangle  \approx 10$, so
this would be an appropriate boundary for the coherent region.  In comparison, the
remainder of the gas spans energies up to about $E= 1500 \hbar \omega$, meaning
that the incoherent region contains many more quantum states and is much larger
than the coherent region.

\section{Equations of motion}\label{eom}
\subsection{Hamiltonian}\label{FT_hamil}

We now substitute the decomposition of equation~(\ref{eqn:fieldsplit})
into the Hamiltonian (\ref{eqn:H}).  
We assume for $k\notin C$ that $\phi_k({\bf x})$ is an eigenstate of
 $\Hsp$, and so $\Ho$
simplifies to
\begin{eqnarray}
\Ho
&=&
\sum_{\p \q}
\langle \p | \Hsp | \q \rangle 
\hat{a}^{\dag}_{ \p}\hat{a}_{\q}
+
\hbar \sum_{k}
\omega_k 
\hat{a}^{\dag}_{ k}\hat{a}_{k}.
\label{eqn:Ho_formalism}
\end{eqnarray}
For the interaction part of the Hamiltonian we have
\begin{mathletters}
\begin{eqnarray}
\Hi 
&=&
\frac{1}{2}
\sum_{\p \q \m \n}
\langle\p \q|V| \m \n \rangle
\hat{a}^{\dag}_{\p}\hat{a}^{\dag}_{\q}
\hat{a}_{\m}\hat{a}_{\n}
\\
&+&
\sum_{\p \q \m n} \left[
\langle\p \q|V| \m n \rangle
\hat{a}^{\dag}_{\p}\hat{a}^{\dag}_{\q}
\hat{a}_{\m}\hat{a}_{n} + h.c \right]
\\
&+&
\frac{1}{2}
\sum_{\p \q mn} \left[
\langle\p \q|V| mn \rangle
\hat{a}^{\dag}_{\p}\hat{a}^{\dag}_{\q}
\hat{a}_{m}\hat{a}_{n} + h.c \right]
\\
&+&
2
\sum_{\p j \m n} 
\langle\p j|V| \m n \rangle
\hat{a}^{\dag}_{\p}\hat{a}_{\m}
\hat{a}^{\dag}_{j}\hat{a}_{n}
\\&+&
2
\sum_{\p j m n} \left[
\langle\p j|V| m n \rangle
\hat{a}^{\dag}_{\p}
\hat{a}^{\dag}_{j}\hat{a}_{m}\hat{a}_{n} + h.c. \right]
\\
&+&
\frac{1}{2}
\sum_{kjmn}
\langle kj|V| mn \rangle
\hat{a}^{\dag}_{k}\hat{a}^{\dag}_{j}
\hat{a}_{m}\hat{a}_{n},
\end{eqnarray}
\label{eqn:Hi_formalism}
\end{mathletters}
\noindent
where  the  symmetrized matrix element 
$\langle kj|V| mn \rangle$ is  defined  in equation~(\ref{eqn:matrix_el}), and $h.c.$
stands for hermitian conjugate.
Using (\ref{eqn:Ho_formalism}) and~(\ref{eqn:Hi_formalism}) we now derive the Heisenberg equations of motion
for the operators in each region.

\subsection{Coherent region}
The evolution of a mode of the coherent part of the field operator  is given by
\begin{mathletters}
\begin{eqnarray}
i \hbar \,\frac{d \hat{a}_{\p}}{d t} &=& 
\sum_{\q}\langle \p |\Hsp | \q \rangle\hat{a}_{\q}  \\
&+&
\sum_{\q\m\n} \langle \p \q |V|\m \n \rangle
\hat{a}_{\q}^{\dag}\hat{a}_{\m}\hat{a}_{\n}
\\
&+&
\sum_{q\m\n}\langle \p q |V|\m \n \rangle \hat{a}_q^{\dag}\hat{a}_{\m}\hat{a}_{\n}\\
&+&
2 \sum_{\q m \n}\langle \p \q |V| m \n \rangle \hat{a}_{\q}^{\dag} \hat{a}_m\hat{a}_{\n}\\
&+&
\sum_{\q mn}\langle \p \q |V|m n \rangle \hat{a}_{\q}^{\dag}\hat{a}_m \hat{a}_n\\
&+&
2 \sum_{qm\n}\langle \p q |V|m \n \rangle
\hat{a}_q^{\dag}\hat{a}_m\hat{a}_{\n}\\
&+&
\sum_{qmn}\langle \p q |V|m n \rangle \hat{a}_q^{\dag}\hat{a}_m\hat{a}_n.
\end{eqnarray}
\label{eqn:acoherent}
\end{mathletters}
in which the coupling to the incoherent region is made explicit.   We now begin
to introduce our approximations.  The condition that a mode is in the coherent
region $C$ is that the population  $\langle \hat{a}_{\nu}^{\dag}
\hat{a}_{\nu}\rangle \gg 1$, which allows us to neglect  the quantum
fluctuations of the projected field operator  $\hat{\psi}({\bf x})$.  Thus we
assume that the region $C$ is well approximated by a mode function given by the
mean value
\begin{eqnarray}
\psi({\bf x}) \equiv \langle \hat{\psi}({\bf x}) \rangle,
\end{eqnarray}
and we can expand the wave function on our basis functions as
\begin{eqnarray}
\psi({\bf x}) &=& \sum_{{\nu} \in C} \langle \hat{a}_{\nu} \rangle
 \phi_{\nu}({\bf x}),
\nonumber\\
&\equiv& \sum_{{\nu} \in C} c_{\nu} \phi_{\nu}({\bf x}).
\label{eqn:condensate_basis}
\end{eqnarray}
\noindent
We obtain the finite temperature GPE by taking the mean value of 
equation~(\ref{eqn:acoherent}).  In this procedure we expand the coherent
region operators as $\hat{a}_{\nu} = c_{\nu} + \hat{\delta}_{\nu}$, and then
neglect the terms involving the quantum fluctuations $\hat{\delta}_{\nu}$.
A typical term simplifies as follows
\begin{eqnarray}
\bigg\langle\sum_{q\m\n}\langle \p q |V|\m \n \rangle 
\hat{a}_q^{\dag}\hat{a}_{\m}\hat{a}_{\n}\bigg\rangle
&\rightarrow&
\sum_{q\m\n}\langle \p q |V|\m \n \rangle 
\langle\hat{a}_q^{\dag}\rangle c_{\m}c_{\n}
\nonumber\\
&=&
\sum_{q}\langle \p q |V|\psi \psi \rangle 
\langle\hat{a}_q^{\dag}\rangle ,
\label{eqn:example_expect}
\end{eqnarray}
where the matrix element is  time dependent as
the wave function
$\psi$ is not a stationary state.  

\begin{figure}\centering
\epsfig{file=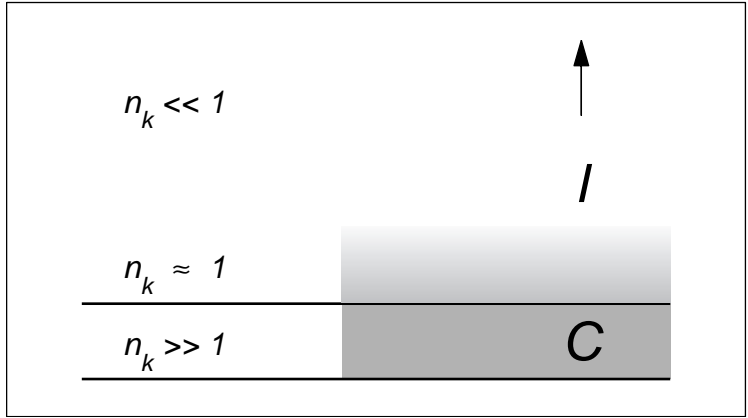}
\caption{
Representation of the degree of coherence in the different regions of the Bose
field.  The shading indicates schematically the coherence of the field.  The
classical region of the field in indicated by $C$, and the incoherent region by 
$I$.  The states in $I$ near the boundary of $C$ will be partially coherent.
}
\label{fig:coherence}
\end{figure}

The incoherent region $I$ is, for the most part, best
represented by number states.  However, this is not always the case.  In
particular the states within $I$ but
near the boundary of the two regions will be partially coherent, as
is illustrated in figure~\ref{fig:coherence}.
The expectation value $\langle\hat{a}_q^{\dag}\rangle$ in this transitional
region will not be zero, and so terms such as (\ref{eqn:example_expect})
are retained in our equations.  This is different from other 
mean field theories in which all coherence is absorbed into the GPE.
In systems that are partially condensed, however, the effect of these terms
 will be small, as the transition region will be narrow compared to the
full width of the incoherent region.

The full equation of motion for a coherent region mode obtained by taking the
mean value of all terms in equation~(\ref{eqn:acoherent}) is
\begin{mathletters}
\begin{eqnarray}
i \hbar \,\frac{d c_{\p}}{d t} &=& 
\langle \p |\Hsp | \psi \rangle\  \\
&+&
 \langle \p \psi |V|\psi \psi \rangle
\label{GPE_usual}
\\
&+&
\sum_{q}\langle \p q |V|\psi \psi \rangle \langle \hat{a}_q^{\dag} \rangle
\label{GPE_adag}
\\
&+&
2 \sum_{ m }\langle \p \psi |V| m \psi \rangle  \langle\hat{a}_m \rangle 
\label{GPE_a}\\
&+&
\sum_{ mn}\langle \p \psi |V|m n \rangle  \langle\hat{a}_m \hat{a}_n \rangle
\label{GPE_aa}\\
&+&
2 \sum_{qm}\langle \p q |V|m \psi \rangle
 \langle\hat{a}_q^{\dag}\hat{a}_m \rangle 
\label{GPE_adaga}
\\
&+&
\sum_{qmn}\langle \p q |V|m n \rangle 
\langle \hat{a}_q^{\dag}\hat{a}_m\hat{a}_n \rangle.
\label{GPE_adagaa}
\end{eqnarray}
\label{eqn:FTGPEbasis}
\end{mathletters}
We can convert this expression to the spatial representation 
by applying the operation $\sum_{\p \in C} |\p\rangle$ to both sides.
Using the contact potential
approximation and recognizing $\sum_{\p \in C} |\p\rangle \langle \p|$ as our
projector of (\ref{eqn:projector}), this procedure
results in an equation we call the 
{\em finite temperature Gross-Pitaevskii equation }
 (FTGPE)
\begin{mathletters}
\begin{eqnarray}
i \hbar \frac{\partial {\psi}({\bf x})}{\partial t}
&=& 
\Hsp \psi({\bf x}) + U_0
\hat{\mathcal{P}}\left\{|\psi({\bf x})|^2 \psi({\bf x})\right\} 
\label{four}
\\
&+&
U_0\hat{\mathcal{P}}\left\{2|\psi({\bf x})|^2
\langle \hat{\eta}({\bf x})\rangle
+
\psi({\bf x})^2 \langle\hat{\eta}^{\dag}({\bf x})\rangle \right\}
\label{three}
\\
&+&
U_0\hat{\mathcal{P}}\left\{\psi^*({\bf x})\langle
\hat{\eta}({\bf x})\hat{\eta}({\bf x})\rangle 
+
2\psi({\bf x})\langle\hat{\eta}^{\dag}({\bf x})\hat{\eta}({\bf x})\rangle \right\}
\label{two}
\\
&+&
U_0\hat{\mathcal{P}}\left\{\langle\hat{\eta}^{\dag}({\bf x})
\hat{\eta}({\bf x})\hat{\eta}({\bf x})\rangle 
\right\},
\label{one}
\end{eqnarray}
\label{eqn:ftgpe}
\end{mathletters}
where $\hat{\eta}({\bf x})$ is defined by (\ref{eqn:eta}).  The FTGPE
constitutes
the main result of this work, and the remainder of this paper is devoted to
discussing the physics of this equation.  

The only approximation that has been made in the derivation of the FTGPE is
that the modes it represents must be highly occupied. No perturbative
techniques have been used, and therefore the equation should  be valid as long
as the condition $N_k \gg 1$ is satisfied.  As a corollary, we expect the FTGPE
could be used to study the region of the phase transition in certain
circumstances.

The FTGPE describes the full dynamics of the region $C$ and its coupling to an
effective thermal bath $\hat{\eta}({\bf x})$.  The initial wave function for the
coherent region will be made up of a sum over a basis with amplitudes with random
phases as is appropriate for a thermal system \cite{kagan3}.  Despite the fact that
the FTGPE is completely unitary and reversible, we  expect that it will evolve
general initial states of $\psi({\bf x})$ to an equilibrium determined by the
temperature and chemical potential of the field $\hat{\eta}({\bf x})$.   This is
because deterministic nonlinear systems exhibit chaotic, and hence ergodic,
behavior if more than a few degrees of freedom are present \cite{riechl}. In fact,
we have shown in reference \cite{ftgpe} that a simplified form of the FTGPE that we
discuss in section \ref{pgpe} can indeed describe evolution towards equilibrium.

Each of the lines of the FTGPE  represents  collision processes involving a
different number of coherent region states.  We describe them briefly here and then
discuss them in more detail in section~\ref{ftgpe_terms}.

The terms on the first line of the FTGPE (\ref{four}) represent purely
coherent region dynamics.  The first term describes the free evolution of the
wave function $\psi$, while the second represents evolution due to two
particles from $C$ colliding, with both particles remaining inside the coherent
region.  

\begin{figure}\centering
\epsfig{file=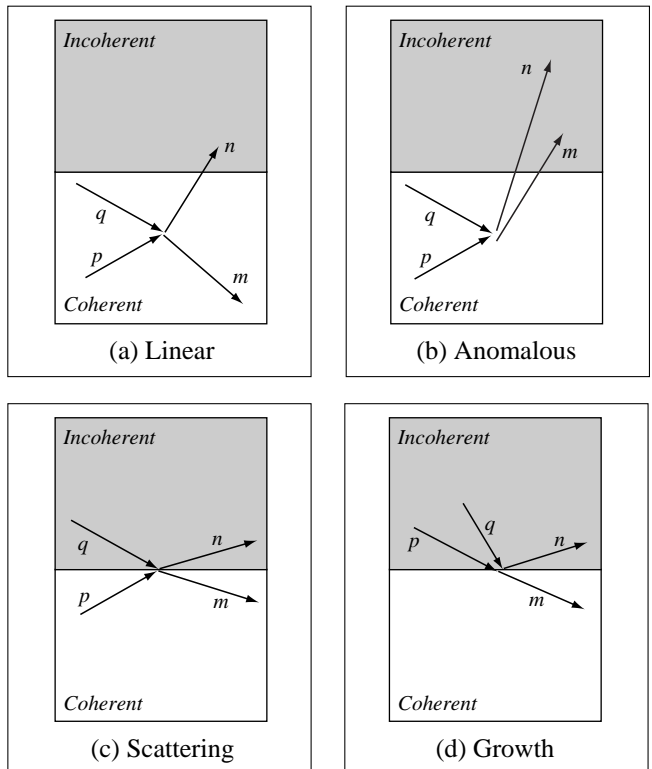}
\caption{An illustrative diagram of the collision processes represented by
 the terms that appear in
the finite temperature GPE (\ref{eqn:ftgpe}) that contain $\eta$ and/or
$\eta^{\dag}$.
(a) The terms linear in $\eta$.  (b) The anomalous term.  (c) The scattering
term.  (d) The growth term. The labels $p$ and $q$ indicate initial states,
and the labels $m$ and $n$ are scattered states.  The reverse processes 
are also possible.}
\label{fig:processes}
\end{figure}

The terms on the second line of the FTGPE (\ref{three}) 
 we refer to as the linear terms. These describe two coherent atoms
interacting, resulting in  one remaining in $C$ and one escaping to the
incoherent region (and the reverse process) and are depicted in
figure~\ref{fig:processes}(a).  These are stimulated processes as the terms
contain  three coherent region labels, and  they  result in the transfer of
some coherence to the bath   $\hat{\eta}({\bf x})$ (see
figure~\ref{fig:coherence}).  However,  because the coherent region is much
smaller than the incoherent region these terms can often be neglected in
comparison with the third and fourth lines. 

The third line of the FTGPE (\ref{two}) will usually be more important than the
second. The first term, which we call the anomalous term,  represents the
collision of two coherent atoms with two incoherent atoms resulting as in
figure~\ref{fig:processes}(b).  If the region $C$ represents only a single
condensate in thermal equilibrium then this term cannot conserve energy and
therefore it cannot describe real processes.\footnote{When the coherent region
is made up of two or more condensates then the first term of 
(\ref{two}) can describe real processes as we discuss in 
section~\ref{sec:etaeta}.} However, it can describe {\em virtual} processes and
thus contributes to the appearance of both the two-body and many-body
T-matrices.  

The second term of the third line of the FTGPE (\ref{two}) represents a
coherent atom colliding with an incoherent atom, with one atom remaining in
each region after the interaction as in figure~\ref{fig:processes}(c).   In
equilibrium this process represents the mean field of the incoherent region
acting on $C$, and  can be added to the  $|\psi({\bf x})|^2$ term of
(\ref{four}).   Away from equilibrium this term  also describes {\em
scattering} processes,  identical to those described in the model of condensate
growth developed in references \cite{BosGro2,QKVI}.

Finally the fourth line (\ref{one}) represents the collision of two
incoherent atoms in which one is transferred to the coherent region $C$
as depicted in figure~\ref{fig:processes}(d).
This is the growth process described in references \cite{BosGro2,QKVI},
and is the main contribution to the transfer of population
between the coherent and incoherent regions.

\subsection{Incoherent region}

The coherence of the majority of levels outside of $C$ is negligible, and
therefore most of the incoherent region is approximately diagonal in the number
state representation. The energy of the quantum levels is large enough that the
mean field of the wave function $\psi$ does not significantly affect this region, and
so we assume that $\Hi$ is a small perturbation to $\Ho$.  Therefore quantum
kinetic theory can accurately describe the evolution of the majority of this
part of the quantum field, with the appropriate modifications to treat the
coupling to the coherent region. 

We can derive an equation of motion for the incoherent region  using similar
techniques to those used in the derivation of the quantum Boltzmann equation
(QBE) and employing two major approximations---the random phase
approximation (RPA) and Wick's theorem \cite{Blaizot}.  
We demonstrate this approach in appendix~\ref{sec:qbe}  where we derive
QBE from the field theory equations.  We also use these methods to simplify the
thermal terms of the FTGPE as we discuss in the next section. 
The kinetic equation for the incoherent region is not the main topic of this
paper; however, for completeness we write it down and discuss the terms that
arise in appendix~\ref{incoherentKE}.

\section{Detailed analysis of the coupling terms of the FTGPE}\label{ftgpe_terms}

In this section we interpret in detail the meaning of, and find expressions for,
the terms involving mean values of combinations of the bath operator 
$\hat{\eta}({\bf x})$ on the right-hand side of the FTGPE,
equation~(\ref{eqn:ftgpe}).   As the incoherent region is  best represented by the
number occupation of the quantum levels, it is appropriate to express these mean
values in terms of the  occupation numbers and to do so we are guided by kinetic
theory.

Expressions are required for the quantities $\langle\hat{\eta}\rangle$, 
$\langle\hat{\eta}^{\dag}\rangle$, $\langle\hat{\eta}\hat{\eta}\rangle$,
$\langle\hat{\eta}^{\dag}\hat{\eta}\rangle$, and 
$\langle\hat{\eta}^{\dag}\hat{\eta}\hat{\eta}\rangle$.  These can be derived
from the equation of motion for the corresponding combinations of creation and
annihilation operators for the incoherent region. The equations we begin with
are rather long, and we write them out
fully in appendix~\ref{a:incoherent}.

We then follow the general procedure used to derive the QBE  (see
appendix~\ref{sec:qbe}). First we approximately integrate the starting equations
for the combinations of creation and annihilation operators, before taking the
mean value and employing Wick's theorem and the RPA to simplify the  resulting
terms such that they depend explicitly on populations $n_p = \langle
\hat{a}_p^{\dag}\hat{a}_p\rangle$.  These expressions are then substituted back
into  $\langle\hat{\eta}\rangle$, 
$\langle\hat{\eta}^{\dag}\rangle$, etc, and then into the FTGPE.  Each of the
terms requires slightly different treatment, and we provide key details in the
following subsections.

\subsection{The linear terms}\label{sec:linear}

The two terms involving $\langle\hat{\eta}({\bf x})\rangle$ or 
$\langle\hat{\eta}^{\dag}({\bf x})\rangle$ in equation~(\ref{eqn:ftgpe})
describe the collision of  two coherent atoms, where one particle  remains in
the region $C$ and  the other is transferred to the incoherent region  (along
with the  reverse process). In systems where there is significant population in
the incoherent region, these terms will not be very large in comparison to the
terms of higher order in $\hat{\eta}$ due to the requirements of energy
conservation and the relatively small size of the region $C$.

Beginning with equation~(\ref{eqn:a2_appendix}), eliminating the  free
evolution via the transformation  $\tilde{a}_p=\hat{a}_p \mbox{e}^{i \omega_p
t}$, taking the mean value, and using Wick's theorem and the RPA, the
only terms that survive are
\begin{eqnarray}
i \hbar \,\frac{d \langle\tilde{a}_p\rangle}{d t} &=& 
\langle p \psi |V|\psi \psi \rangle
\mbox{e}^{i\omega_p t}
+2 \sum_{q}\langle p q |V|q \psi \rangle n_q 
\mbox{e}^{i\omega_p t}.
\end{eqnarray}
We can also drop the last term in this equation using the rotating wave
approximation, as the components of  $\psi$ will be oscillating at frequencies
smaller than $\omega_p$.
Thus the result is
\begin{eqnarray}
i \hbar \,\frac{d \langle\tilde{a}_p\rangle}{d t} &=& \langle p \psi |V|\psi \psi \rangle
\mbox{e}^{i\omega_p t}.
\label{agpe}
\end{eqnarray}

In order to find a approximate solution for equation~(\ref{agpe}), we assume that we
can expand the coherent region wave function in a basis that is approximately
diagonal---essentially a quasiparticle basis.  We write
\begin{equation}
\psi({\bf x}) = \sum_{\n} \tilde{c}_{\n} \xi_{\n}({\bf x}) \mbox{e}^{-i\omega_{\n} t},
\label{eqn:psi_basis_diag}
\end{equation}
where the coefficients $\tilde{c}_{\n} = c_{\n} \mbox{e}^{i \omega_{\n} t}$. 
The equation of motion for $\tilde{a}_p$ becomes
\begin{eqnarray}
i \hbar \,\frac{d \langle\tilde{a}_p\rangle}{d t} &=& 
\sum_{\q\m\n}
\tilde{c}_{\q}^* \tilde{c}_{\m} \tilde{c}_{\n} 
\langle p \q|V |\m \n \rangle
\mbox{e}^{i\omega_{p\q\m\n}t},
\label{eqn:notime}
\end{eqnarray}
where we have introduced the notation
\begin{equation}
\omega_{pqmn} \equiv \omega_p + \omega_q - \omega_m -\omega_n.
\label{eqn:omega4}
\end{equation}
If the set $\{\xi_{\n}\}$ is a good basis for the coherent region, then the exponential term 
contains most of the time dependence of equation~(\ref{eqn:notime}).
We can therefore take everything else outside the integral, and use the standard
result
\begin{equation}
\frac{1}{x + i\epsilon} = P\left(\frac{1}{x}\right) - i \pi \delta(x),
\end{equation}
to find an approximate solution to equation~(\ref{eqn:notime}).
Incorporating the free
evolution in the solution, we find
\begin{eqnarray}
\langle\hat{a}_p\rangle &=& \sum_{\q\m\n} \bigg\{\frac{
{c}_{\q}^* {c}_{\m} {c}_{\n}
\langle p \q|V |\m \n \rangle}
{\hbar \omega_{p\q\m\n}}
- i\pi {c}_{\q}^* {c}_{\m} {c}_{\n}
\langle p \q|V |\m \n \rangle
\delta(\hbar\omega_{p\q\m\n})
\bigg\}.
\label{eqn:a_soln}
\end{eqnarray}
Since we are mainly interested in the kinetic processes that can occur,
we neglect the energy shift described by the principal part (the first term
on the right-hand side).
The relevant term in the FTGPE in basis form (\ref{GPE_a}) is thus
\begin{eqnarray}
i \hbar \,\frac{d c_{\p}}{d t} &=& ...
-2 \pi i \sum_{ p \x \y }\langle \p \x |V| p \y \rangle c_{\x}^* c_{\y}
\sum_{\q\m\n} 
 {c}_{\q}^* {c}_{\m} {c}_{\n}
\langle p \q|V |\m \n \rangle
\delta(\hbar\omega_{p\q\m\n}),
\label{eqn:GPEa}
\end{eqnarray}
where we have expanded all the condensate wave functions as in 
(\ref{eqn:psi_basis_diag}). We note that equation~(\ref{eqn:GPEa})
contains only coherent region amplitudes $\{c_{\p}\}$ because the processes we
have included are all stimulated.

A situation where the terms discussed in this section may become important is
for experiments in which a Bose condensate near $T=0$ is disturbed by a sudden
change in its scattering length by the use of a Feshbach resonance.  Such
experiments have been carried out by Donley {\em et al.}\ using $^{85}$Rb at
JILA \cite{Cornish}.   The change in scattering length causes the collision
processes represented by the linear terms of the FTGPE to become energetically
allowed.  This description of the physics is identical to the argument by Duine
and Stoof that the loss observed can be explained via elastic collisions due to
an imaginary part of the many-body T-matrix \cite{Duine}.  We will address this
issue further in a future paper.

\subsection{The anomalous term}\label{sec:etaeta}

We now consider the term involving $\langle \hat{\eta} \hat{\eta} \rangle$ on
the third line of the FTGPE (\ref{two}). 
Expanding this quantity in the incoherent region basis gives
\begin{eqnarray}
\langle \hat{\eta} \hat{\eta} \rangle 
&=& \sum_{mn} \phi_m \phi_n\langle
\tilde{a}_m \tilde{a}_n \rangle e^{-i(\omega_m + \omega_n)t}.
\end{eqnarray}
To find an expression for $\langle
\tilde{a}_m \tilde{a}_n \rangle$ we use the equation of motion for this
quantity given as equation~(\ref{eqn:aa}).
Eliminating the free evolution, taking the mean value,
expanding the coherent region
wave function in the approximately diagonal basis according to
equation~(\ref{eqn:psi_basis_diag}), and finally 
using Wick's theorem and the RPA we obtain
\begin{eqnarray}
i \hbar \frac{\partial \langle \tilde{a}_m \tilde{a}_n\rangle}{\partial t} 
&=& 
 \sum_{\m \n}\tilde{c}_{\m} \tilde{c}_{\n}
 \langle mn  |V|\m \n \rangle \mbox{e}^{i\omega_{mn\m\n}t}
 (1 + n_m + n_n)\nonumber\\
 &&
 + \sum_{kj} \langle mn |V|kj \rangle \langle\tilde{a}_k \tilde{a}_j 
\rangle e^{i \omega_{mnkj}t}(1 + n_m + n_n).
\label{eqn:aa_expand}
\end{eqnarray}
The first line of this equation describes two particles  scattering from  the
coherent region into states $m$ and $n$. The second line  would usually be
ignored in the RPA, as it is of higher order than the first line.  However the
matrix element of this line describes particles from $(m,n)$ scattering
into $(k,j)$ and then onto other states.  This offers the possibility of ladder
diagrams, such that the two particles could scatter back into the coherent
region without interacting with a third atom.  We retain this term in
equation~(\ref{eqn:aa_expand}) as the mean value it contains is  of the same
form as the that on the left-hand side. Because of this,
equation~(\ref{eqn:aa_expand}) has the form of a  Lippmann-Schwinger equation
in the time domain, and we will see that this is  where the T-matrix appears in
the formalism.  

To solve this equation we start by considering the first line only, as it is
the lowest order term.   Once again, most of the time dependence is contained
within the exponential, and so  we find the solution is 
\begin{eqnarray}
\langle\tilde{a}_m \tilde{a}_n\rangle 
&=& 
\sum_{\m \n}\tilde{c}_{\m}
\tilde{c}_{\n}\frac{\langle mn  |V|\m \n \rangle  (1 + n_m +
n_n)e^{i\omega_{mn\m\n}t}} {\epsilon_{\m} + \epsilon_{\n} - \epsilon_m -
\epsilon_n}  \nonumber\\ && - i \pi \sum_{\m \n}\tilde{c}_{\m}
\tilde{c}_{\n}\langle mn  |V|\m \n \rangle \delta(\epsilon_{\m} + \epsilon_{\n} -
\epsilon_m - \epsilon_n). 
\label{eqn:aa_soln1} 
\end{eqnarray} 
For a single condensate near thermal equilibrium, the energy delta-function can
never be satisfied as it requires two low-energy, coherent atoms from within the
coherent region to collide and result in two high-energy, incoherent atoms.  We will
return to this point later in the section.
We assume that the full
solution of equation~(\ref{eqn:aa_expand}) is of the same form as 
(\ref{eqn:aa_soln1}) but with the interaction potential $V$ replaced by a
T-matrix $T^{\rm up}$
\begin{eqnarray}
\langle \hat{a}_m \hat{a}_n\rangle
&=&
\sum_{\m\n}{c}_{\m} {c}_{\n}\frac{\langle mn  |T^{\rm up}|\m \n \rangle}
{\epsilon_{\m} + \epsilon_{\n} - \epsilon_m - \epsilon_n}.
\label{aa:soln}
\end{eqnarray}
This is a solution of equation~(\ref{eqn:aa_expand}) if the operator $T^{\rm up}$ obeys
\begin{eqnarray}
T^{\rm up}(z) &=& V + \sum_{kj}V |kj\rangle \frac{1 + n_k +n_j}
{z - \epsilon_m - \epsilon_n}
\langle kj | T^{\rm up}(z),
\label{its_a_tmatrix}
\end{eqnarray}
where we have identified the parameter $z = \epsilon_{\m} + \epsilon_{\n} +
i\delta$ as the incoming energy of the two particles in the collision.   The 
small imaginary part $i\delta$ in this parameter generates  the delta function
term in (\ref{eqn:aa_soln1}). Equation~(\ref{its_a_tmatrix}) is the
definition of a {\em restricted} many-body T-matrix as the indices $k,j$ are defined
to be outside the coherent region. 

The many-body T-matrix describes collisions in the presence of a medium.  It
takes into account the fact that the virtual states that two particles pass
through in the collision may be occupied, and the scattering rate enhanced by a
factor of $(1 + n_k + n_j)$.  However, as the states $k,j$ are in the
incoherent region the populations are generally small, and if we can
approximate $n_k = n_j = 0$ we recover the definition of a restricted two-body
T-matrix.  

If we now substitute the solution equation~(\ref{aa:soln}) into the basis set FTGPE
equation~(\ref{eqn:FTGPEbasis}), we find from the terms (\ref{GPE_usual}) and 
(\ref{GPE_aa})
\begin{eqnarray}
i \hbar \,\frac{d c_{\p}}{d t} &=& \ldots +
\sum_{\m \n} c_{\m} c_{\n} \langle \p \psi |V|\m \n \rangle+
\sum_{ mn}\langle \p \psi |V|m n \rangle  \sum_{\m\n}{c}_{\m} {c}_{\n}\frac{\langle mn  |T^{\rm up}|\m \n \rangle}
{\epsilon_{\m} + \epsilon_{\n} - \epsilon_m - \epsilon_n},\nonumber\\
&=&  \sum_{\m \n} c_{\m} c_{\n} \langle \p \psi |
\left[V + 
\sum_{ mn}V|m n \rangle \frac{\langle mn  |T^{\rm up}}
{\epsilon_{\m} + \epsilon_{\n} - \epsilon_m - \epsilon_n} \right]
|\m \n \rangle,\nonumber\\
&\equiv& \langle \p \psi | T^{\rm up}|\psi \psi\rangle.
\end{eqnarray}
Thus the anomalous term introduces the restricted T-matrix into coherent region
collisions, and it is appropriate to approximate this T-matrix by a contact
potential \cite{sam}.  With this treatment of the anomalous term we have
carried out the ultra-violet renormalization of the theory.  

It is useful to make two additional points. First, the T-matrix that
enters our equations is not the full two-body T-matrix as the sum over
intermediate states only includes levels in the incoherent region.  Thus the
contact potential we use in our calculations should be
\begin{equation}
T^{\rm up} \rightarrow \tilde{U}_0 \delta({\bf x} - {\bf x}'),
\end{equation}
where $\tilde{U}_0 \neq U_0$.
The remainder of the terms
that would `upgrade' $\tilde{U}_0$ to $U_0$ are  included
directly in the simulation of the coherent region using the FTGPE, and so all
T-matrix effects are actually included in the formalism.

In practice we find these two quantites are related in the
homogeneous limit \cite{sam} by
\begin{equation}
\tilde{U}_0  =  \frac{U_0}{ 1 - U_0 \alpha_K},
\end{equation}
where $\alpha_K$ is defined by
\begin{equation}
\alpha_K = \frac{1}{(2\pi)^3} \int_0^K d^3 {\bf k} \frac{m}{\hbar^2 k}
= \frac{mK}{2 \pi^2 \hbar^2},
\end{equation}
and $K$ is the wave vector of the cutoff between the coherent and incoherent
region.  Thus as long as the condition 
$U_0 \alpha_K \ll 1$, or equivalently $Ka \ll 1 $ is satisfied,  we are
justified in setting $\tilde{U}_0 \approx U_0$.  As in any calculation the
coherent region will be rather small, and experimentally measured
scattering lengths for alkali atoms are generally not known with an accuracy of
better than 10\%, this approximation is well justified in the homogeneous case. 
It seems reasonable to expect the same condition should hold true in a trap.

The second point is that the remaining terms of the FTGPE still have the
interatomic potential rather than the T-matrix in their matrix elements.  It
turns out that it is reasonable to use the contact potential approximation in
these, although we have not explicitly upgraded the matrix elements to
T-matrices.  For further details we refer the reader to
references \cite{martin,sam,prou,Stoof_Tmatrix}

\subsubsection{Elastic loss in condensate collisions}

In section~\ref{sec:etaeta} we stated that the delta function in the solution of the anomalous 
term (\ref{eqn:aa_soln1}) could not be satisfied for a system with only one
condensate near thermal equilibrium.  The situation is different when we
consider the collision of two condensates, where the T-matrix can have an
imaginary part.

In the formalism described here the coherent region $C$ is defined such that
it contains only the modes of the gas whose occupation number satisfies
$N_k \gg 1$.  If we consider a condensate that has been formed in a harmonic
trap, but then quickly released into free space, we can analyse the wave
function in terms of a plane-wave basis.  The region $C$ will typically be
defined by a small spherical or ellipsoidal region in $k$-space about a central
wave-vector ${\bf K}$.  For a condensate released at rest  we have  ${\bf K} =
0$, but by applying a Bragg-pulse to the condensate  on release, it can be split
into two equal parts---one at rest and one with a momentum  $\hbar {\bf K}$ 
\cite{Phillips_bragg}.  In the centre-of-mass frame, the two condensates
have momenta $-\hbar {\bf K}/2$ and $+\hbar {\bf K}/2$ respectively.

\begin{figure}\centering
\epsfig{file=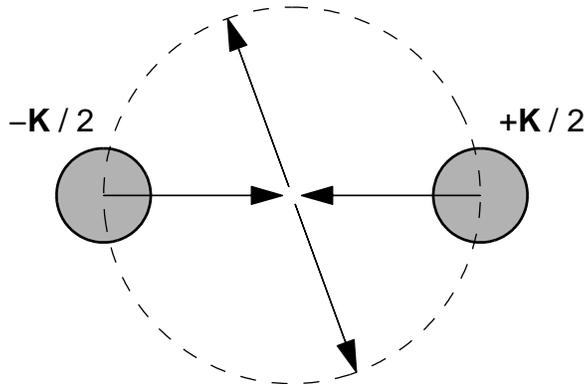}
\caption{A depiction in {\em k}-space of two condensates colliding 
in the centre-of-mass frame.  The two
shaded areas  centred about $\pm {\bf K}/2$ indicate the coherent region $C$. 
The arrows indicate a possible collision process between two coherent particles, in
which both end up outside  the coherent region.  The  dashed circle
indicates the region of all possible collisions that conserve both energy 
and momentum.}
\label{fig:collision}
\end{figure}

Now consider this system analysed in the plane-wave basis. If $\hbar {\bf K}$
is large compared to the momentum width of the two condensates, then the region
$C$ will be made up of two distinct parts of {\em k}-space, as depicted in
figure~\ref{fig:collision}. This means that
in the collision of the two condensates it is possible for an atom from each to
collide, and then scatter into any spatial direction while energy is still
conserved.  A large number of these collisions will take both particles outside
the region $C$, as depicted in figure~\ref{fig:collision}.  As these processes
cause  scattering of both particles  into modes that are otherwise empty (and
are hence spontaneous), they cannot be represented by the GPE 
(\ref{eqn:gpe}) (see appendix \ref{GPEkinetic} for further explanation).
However, these collisions can be represented in the FTGPE  via the anomalous
term  $\langle \hat{\eta} \hat{\eta}\rangle$.  Unlike the case of a single
condensate, the delta function part of  equation~(\ref{eqn:aa_soln1}) can now be
satisfied, and therefore real transitions can occur.  

We can also see that if the relative momenta of the two condensates is not
large, then the two parts of the region $C$ will overlap and most of the
circumference of the  dashed circle in figure~\ref{fig:collision} will lie within
$C$.  This means that the GPE can describe condensate collisions at low momenta
for which spontaneous collisions can be neglected. However, at high relative
momenta other methods, such as described here, are required.

The process of elastic loss is a form of spontaneous Beliaev damping.  An
analogous phenomenon for a trapped BEC is, for example, when a high energy
coherent collective excitation is generated in a ground state condensate.
The coherent region is again divided into two parts, and the excitation can
interact with the ground state condensate and spontaneously decay into two
lower energy quasiparticles.

\subsubsection{Four-wave mixing}

In an elegant experiment, the Phillips group at NIST demonstrated the
atom-optical equivalent of four-wave mixing with Bose-Einstein condensates 
\cite{Phillips_4wave}. After releasing a condensate  from a trap, two separate
Bragg pulses were applied in succession such that the BEC split into three
distinct parts, each with a different momentum (two moving and one at rest in
the lab frame).  For appropriately chosen momentum values, a fourth condensate
was observed to appear. 

This experiment can be understood by considering figure~\ref{fig:collision}, but
now with a third condensate at the tip of the arrow at the top of the dashed
circle.  While the entire dashed region is still available in collisions
between atoms from the first two condensates, the presence of third condensate
stimulates transitions into this particular mode, resulting in the formation of
a condensate at the tip of the downward pointing arrow. 

The ordinary GPE (\ref{eqn:gpe}) can describe the formation of the fourth
condensate as this is a stimulated collision process
\cite{4wavetheory1,4wavetheory2}. However, the new condensate that appears is
entangled with the atoms that are scattered into the third condensate as the
colliding atoms are necessarily correlated.  This correlation {\em cannot} be
described by the ordinary GPE, and other methods are required 
\cite{4wave_entangle}.  We emphasize that such effects can be
described by the anomalous term in the FTGPE.

\subsection{The scattering term}
We now consider the term containing $\langle \hat{\eta}^{\dag}
\hat{\eta}\rangle$ in the third line of the FTGPE (\ref{two}).  To find an
explicit expression for this quantity, we
first eliminate the free evolution of equation~(\ref{eqn:adaga_appendix}), before
taking the mean value and making  use of Wick's theorem and the
 RPA to find the equation of motion
\begin{equation}
i \hbar \frac{\partial \langle \tilde{a}_m^{\dag} \tilde{a}_n \rangle}{\partial t}
=
2 \sum_{\m \n}
\tilde{c}_{\m}^* \tilde{c}_{\n}
\langle   \m m|V| \n n\rangle (n_m - n_n) 
e^{i \omega_{\m m \n n} t}.
\label{eqn:adaga_solve}
\end{equation}
The solution of equation~(\ref{eqn:adaga_solve}) differs slightly from that of 
the anomalous and linear terms  we 
considered earlier.  We previously made the assumption that the approximate
solutions were zero at time $t=-\infty$, but this is not the case here.
The diagonal term ($m=n$) of $\langle \tilde{a}_m^{\dag} \tilde{a}_n \rangle$
is the average  number of particles in mode $n$. 
For a system at finite temperature in which the incoherent region begins with
some population, this will have a non-zero initial value, and
appears in the solution
of equation~(\ref{eqn:adaga_solve}) as a constant of integration. We find
\begin{equation}
\langle \hat{a}_m^{\dag} \hat{a}_n \rangle
= -2 \pi i \sum_{\m \n}{c}_{\m}^* {c}_{\n}
\langle   \m m|V| \n n\rangle (n_m - n_n) 
\delta(\hbar \omega_{\m m \n n}) + n_m\delta_{mn},
\end{equation}
where in the spatial representation of the FTGPE
the term involving the Kronecker delta function
represents the mean field of the incoherent region acting on the coherent region wave function.
We have neglected the principal part in the solution 
as it is dominated by the energy conserving processes.

In the basis set representation of the FTGPE the corresponding term 
(\ref{GPE_adaga}) is therefore
\begin{eqnarray}
i \hbar \,\frac{d c_{\p}}{d t} &=& \ldots  +
2 \sum_{m \q} n_m c_{\q}\langle \p m |V|m \q \rangle\\
&-&4\pi i \sum_{mn \q} c_{\q}\langle \p m |V|n \q \rangle
\sum_{\m \n}{c}_{\m}^* {c}_{\n}
\langle   \m m|V| \n n\rangle (n_m - n_n) 
\delta(\hbar \omega_{\m m \n \n}) \nonumber
\end{eqnarray}
The processes it describes are analogous to the scattering term that was
introduced in the model of condensate growth in references \cite{BosGro2,QKVI}.  It
represents an incoherent particle colliding with a coherent particle, with the
coherent particle moving between levels within the region $C$.  In the presence
of a condensate this process is recognized as Landau damping, and in 
references \cite{BosGro2,QKVI}  it was shown to have an important effect  in the
onset of condensate growth.  However in simulations closer to equilibrium when
a condensate is already present, the off-diagonal part of this term is unlikely
to be large as the forward and backward rates will be similar.

\subsection{The growth term}

Finally, we consider the term involving  $\langle \hat{\eta}^{\dag} \hat{\eta}
\hat{\eta} \rangle$ on the fourth line of the FTGPE (\ref{one}) which
we identify as the growth term.  This will generally be the most important term
involving the  field $\hat{\eta}$ in the FTGPE, as it will be responsible for
the majority of particle exchange between the coherent and incoherent regions. 
While the linear terms and the anomalous terms can be important in certain
circumstances near $T=0$, in most situations at finite temperature they are
small in comparison with the growth term because of the size difference between
the coherent and incoherent regions.

To find an explicit expression for this term we begin with the equation of
motion for $\hat{a}^{\dag}_q \hat{a}_m \hat{a}_n$,
 equation~(\ref{eqn:adagaa_appendix}).   Eliminating the free evolution, taking the
 mean value and  making use of Wick's
theorem and the RPA leaves us with
\begin{eqnarray}
i \hbar
\frac{\partial
\langle \tilde{a}^{\dag}_q \tilde{a}_m \tilde{a}_n \rangle}{\partial t}
&=& 
2\sum_{\m}\tilde{c}{\m}\langle mn |V| q \m \rangle e^{i\omega_{mnq\m}t}
 \{n_q(1 + n_m + n_n) - n_m n_n\}.
\end{eqnarray}
As this can represent energy conserving processes, the approximate solution is
\begin{eqnarray}
\hspace*{-10mm}\langle \tilde{a}^{\dag}_q \tilde{a}_m \tilde{a}_n \rangle
&=& 
 2 \pi i\sum_{\m}\tilde{c}_{\m}\langle mn |V| q \m \rangle \delta(\hbar \omega_{mnq\m})
 \{n_m n_n - n_q(1 + n_m + n_n)\}.
\label{eqn:adagaa_soln1}
\end{eqnarray}
Substituting equation~(\ref{eqn:adagaa_soln1}) into the basis set version 
of the FTGPE we find the term corresponding to equation~(\ref{GPE_adagaa}) becomes
\begin{eqnarray}
i \hbar \,\frac{d c_{\p}}{d t} &=& \ldots +
 2 \pi i \sum_{qmn\m}{c}_{\m}\langle \p q |V|m n \rangle 
\langle mn |V| q \m \rangle \delta(\hbar \omega_{mnq\m})
\nonumber \\ && \times
 \{n_m n_n - n_q(1 + n_m + n_n)\}.
\label{GPE_growth_basis}
\end{eqnarray}
We can understand the physical meaning of this result by comparing it with the
equivalent quantum Boltzmann rate for the growth of a coherent region level with
population $N_{\p}$ (see equation \ref{eqn:QBE}).  Making the approximation that $(1 + N_{\p}) \approx 
N_{\p}$, we find
\begin{eqnarray}
\frac{d N_{\p}}{dt} &\propto& 
 \sum_{qmn} 
(1+N_{\p})(1+n_q)n_m n_n - N_{\p} n_q (1 +n_m) (1+n_n),\nonumber\\
&\approx&
\sum_{qmn} N_{\p} [n_m n_n - n_q( 1 + n_m + n_n)],
\label{eqn:growth_ke_bit}
\end{eqnarray}
and we can see that the right-hand sides of 
(\ref{GPE_growth_basis}) and~(\ref{eqn:growth_ke_bit}) are similar.

Equation~(\ref{GPE_growth_basis}) is very similar to the growth part of the
description of condensate formation of reference \cite{BosGro2}.  However, rather
than describing the rate of change of an occupation number of a quasiparticle
level, it describes the growth in amplitude of a  basis component making up the
coherent region  wave function $\psi$.  To calculate this term for inclusion in
the FTGPE requires both a reasonably good basis and corresponding energies  for
the coherent region, along with a method of calculating or approximating  the
matrix elements that appear in equation~(\ref{GPE_growth_basis}). While this is not
difficult in principle, in practice they need to be calculated at each time
step in the evolution of the FTGPE.  This issue will be treated in future
numerical work.

\section{The projected Gross-Pitaevskii equation}\label{pgpe}

The finite temperature Gross-Pitaevskii equation contains all the necessary
elements for a complete description of a condensed Bose gas, given that the
occupation number conditions are satisfied.  However, it is somewhat
complicated to implement numerically.    Some insight can be gained by
neglecting all the terms coupling the coherent region to the effective bath
$\hat{\eta}$, and considering the equation of motion for
the coherent region region alone---the first line of equation~(\ref{eqn:ftgpe}).  We
call this the projected Gross-Pitaevskii equation (PGPE)
\begin{equation}
i \hbar \frac{\partial {\psi}({\bf x})}{\partial t} = 
\Hsp \psi({\bf x}) + U_0
\hat{\mathcal{P}}\left\{|\psi({\bf x})|^2 \psi({\bf x})\right\},
\label{eqn:pgpe}
\end{equation}
and we have studied solutions of this in reference \cite{ftgpe}.
Although it is not immediately evident, the PGPE does conserve normalization and
energy, and this can be easily understood by considering the effective Hamiltonian as
we discuss further in appendix~\ref{PGPEnorm}.

The projected GPE describes a microcanonical system, whereas in the full FTGPE
the region $C$ will fluctuate in energy and number of particles.  However, if
the region $C$  contains sufficiently many modes, then fluctuations in energy
and particle number in the grand canonical ensemble would be small.  Hence we
would expect an equilibrium state of the projected GPE to be similar to
that of the finite temperature GPE coupled to a bath $\hat{\eta}({\bf x})$ with
the appropriate chemical potential and temperature. 

The projected GPE by itself cannot capture the entire physics of the Bose field
at finite temperature. Indeed, in any system in which there is a significant
thermalized coherent region that may be modelled by this equation, there will be
a much larger incoherent region whose effects will be important.  Nonetheless,
we showed in reference \cite{ftgpe} that the GPE {\em without any additional terms}
can describe evolution of general configurations of the coherent region $C$
towards an equilibrium that can be parameterized by a temperature.

The detailed non-equilibrium dynamics of the system {\em will} depend on the
exchange of energy and particles between $C$ and the bath. However, we expect
that for modelling  many experiments at finite temperature the projected GPE
alone may be sufficient.  It is well known from kinetic theory that particles
mainly interact with others of similar energy.  If the presence of the bath is
important, then it could be well approximated by a constant temperature and
chemical potential, along the lines of the formalism outlined in
reference \cite{QKIII} and used in the model of condensate growth in
references \cite{BosGro2,QKVI}.  This issue will be considered further elsewhere.

\section{Conclusions and prospects}\label{conclusions}

In this paper we have derived an approximate formalism for calculating the
dynamics of a thermal Bose gas in the highly-occupied limit. We have derived a
non-perturbative finite temperature Gross-Pitaevskii equation that describes the
evolution of the wave function of the coherent region $C$, and identified and
discussed the physical meaning of each of the terms that arise.  In particular
we have  indicated how the anomalous term introduces the T-matrix in coherent
region collisions, and that this can describe elastic particle loss in
condensate collisions at large relative momenta.  The terms analogous to the 
scattering and growth processes of reference \cite{BosGro2} have also been
discussed. 

While the formalism discussed here leaves out the possibility of quantum
correlation effects in BECs, it should nevertheless provide a basis for the
description of many experimentally accessible features that can not be described
by the ordinary Gross-Pitaevskii equation. An important aim of our development
has been to produce a numerically tractable formalism, and the first step in a
practical implementation has been made in reference \cite{ftgpe} where it was shown
that the projected GPE will evolve a generalized random initial distribution to
an equilibrium described by a temperature. The next step is to include some of
the additional terms of the FTGPE described in section~\ref{ftgpe_terms}.

We can also expect the FTGPE to describe the phase transition region, as long as
the condition $N_k \gg 1$ is satisfied for the coherent region modes.  The
physics of phase transitions is generally classical in nature, being dominated
by large fluctuations at long wavelengths. This of course is what the GPE
describes, and in fact it has been used as a model of phase transitions in other
areas of condensed matter physics. The GPE has the same energy functional as
that used in the classical renormalization group theory of the superfluid phase
transition, and it is therefore reasonable to expect  the same approximations to be valid for
the case of BEC. Our formalism, which couples the GPE to a kinetic treatment of
the thermal particles, provides a non-perturbative dynamical finite temperature
theory of BEC. This can then be used to study the region of the phase transition
where perturbative approaches fail, and indeed where no other techniques are
available.

\ack
This work was financially supported by the UK EPSRC.  We thank Dr. Sam Morgan
for many useful discussions, suggestions, and for proof-reading the manuscript.
MJD would like to acknowledge St John's College, Oxford for additional financial
support.

\begin{appendix}

\section{Derivation of the quantum Boltzmann equation}\label{sec:qbe}

In this appendix we give a derivation of the quantum Boltzmann equation, which
gives an accurate description of the time evolution of a Bose gas well above
the transition temperature.  In this regime the mean time between particle
collisions is long compared to the duration of a collision,  so the 
eigenstates of  $\Ho$ provide a good basis and the interaction
part of the Hamiltonian $\Hi$ can be treated as a perturbation.  This is the
method by which incoherent region terms are treated in this paper.

The operators $\tilde{a}_p$ have no mean value above the transition temperature,
and so we want an equation of motion for the mean number of
particles in mode $p$, $\langle \hat{n}_p \rangle = \langle
\tilde{a}_p^{\dag}\tilde{a}_p\rangle$. From equation~(\ref{eqn:at}) we find
\begin{equation}
\frac{d \hat{n}_p}{dt} = 
-\frac{i}{\hbar} \sum_{qmn}\langle pq|V|mn\rangle 
\tilde{a}_p^{\dag}\tilde{a}_q^{\dag}\tilde{a}_m\tilde{a}_n
e^{i(\omega_p + \omega_q - \omega_m - \omega_n)t} + h.c. ,
\label{eqn:n}
\end{equation} 
where $h.c.$ is the hermitian conjugate.  In order to  
find a closed expression for the evolution of $\hat{n}_p$, we begin by finding
 an equation of motion
for the quantity $\tilde{a}_p^{\dag}\tilde{a}_q^{\dag}\tilde{a}_m\tilde{a}_n$
which appears 
on the right-hand side of equation~(\ref{eqn:n}).  
However,  this new equation for four operators contains terms 
involving six operators, and the equations of
motion for six operators involve eight operators and so on.  We proceed by
truncating this series at the first iteration, approximately solving the equation
for $\tilde{a}_p^{\dag}\tilde{a}_q^{\dag}\tilde{a}_m\tilde{a}_n$, and substituting
the result back into equation~(\ref{eqn:n}).

 One way to derive an equation of motion for 
$\tilde{a}_p^{\dag}\tilde{a}_q^{\dag}\tilde{a}_m\tilde{a}_n$ 
would be to commute it with the Hamiltonian.  An equivalent method
(which will be useful for other purposes later) is simply to use the chain rule.
We can formally write the solution as
\begin{eqnarray}
\tilde{a}_p^{\dag}\tilde{a}_q^{\dag}\tilde{a}_m\tilde{a}_n
& = &
\int_{-\infty}^{t}dt' 
\frac{d}{dt'}\left(
\tilde{a}_p^{\dag}\tilde{a}_q^{\dag}\tilde{a}_m\tilde{a}_n
\right), \nonumber \\
&=& \int_{-\infty}^{t}dt' 
\left[
  \frac{d\tilde{a}_p^{\dag}}{dt'}\tilde{a}_q^{\dag}\tilde{a}_m\tilde{a}_n
+ \tilde{a}_p^{\dag}\frac{d\tilde{a}_q^{\dag}}{dt'}\tilde{a}_m\tilde{a}_n
+ \tilde{a}_p^{\dag}\tilde{a}_q^{\dag}\frac{d\tilde{a}_m}{dt'}\tilde{a}_n
+ \tilde{a}_p^{\dag}\tilde{a}_q^{\dag}\tilde{a}_m\frac{d\tilde{a}_n}{dt'}
\right].
\nonumber\\
\label{eqn:big_n_long}
\end{eqnarray}
We now substitute equation~(\ref{eqn:at}) for each of the $d\tilde{a}_k/dt$. 
This is a straight-forward but tedious process, which we illustrate only 
on the last term of equation~(\ref{eqn:big_n_long}). We have
\begin{eqnarray}
\lefteqn{\int_{-\infty}^{t}dt' 
\left(
\tilde{a}_p^{\dag}\tilde{a}_q^{\dag}\tilde{a}_m\frac{d\tilde{a}_n}{dt'}
\right)}
\nonumber\\
&=&-\frac{i}{\hbar}\sum_{jkl}\delta_{in} \langle ij|V|kl\rangle
\int_{-\infty}^{t}dt'\tilde{a}_p^{\dag}\tilde{a}_q^{\dag}\tilde{a}_m 
\tilde{a}_j^{\dag}\tilde{a}_k\tilde{a}_l
e^{i(\omega_i + \omega_j - \omega_k - \omega_l)t'},
\nonumber\\
&=&-\frac{i}{\hbar}\sum_{jkl}\delta_{in} \langle ij|V|kl\rangle
\int_{-\infty}^{t}dt'\nonumber\\
&&\times \mbox{} 
(\tilde{a}_p^{\dag}\tilde{a}_q^{\dag}\tilde{a}_j^{\dag}
\tilde{a}_m\tilde{a}_k\tilde{a}_l
+
\tilde{a}_p^{\dag}\tilde{a}_q^{\dag}\tilde{a}_k\tilde{a}_l\delta_{mj})
e^{i(\omega_i + \omega_j - \omega_k - \omega_l)t'}.
\end{eqnarray}
where we have arranged the creation and annihilation operators
in normal order.

We now introduce our first approximation.   The free evolution of the operators 
has already been removed and so assuming the interaction is a perturbation, over
the period of the integral most of the time dependence is contained in the
exponential.   Invoking the Markov approximation (which assumes that
correlations between the operators are not important on  the time scale of
interest), we can therefore take the operators outside of the integral and
replace their time dependence by the time at the upper limit.  
 The integral then leaves us with 
a delta function when the
frequencies of the modes add to zero, and a principal part when they do not---a
standard result normally expressed as
\begin{equation}
\frac{1}{x + i\epsilon} = P\left(\frac{1}{x}\right) - i \pi \delta(x).
\end{equation}
We assume that the principal part is negligible, which leaves us with
\begin{eqnarray}
\int_{-\infty}^{t}dt' 
\left(
\tilde{a}_p^{\dag}\tilde{a}_q^{\dag}\tilde{a}_m\frac{d\tilde{a}_n}{dt'}
\right)
&=&-\frac{i\pi }{\hbar}\sum_{jkl}\delta_{in} \langle ij|V|kl\rangle
\delta(\omega_i + \omega_j - \omega_k - \omega_l)
\nonumber\\
&& \times \mbox{}(\tilde{a}_p^{\dag}\tilde{a}_q^{\dag}\tilde{a}_j^{\dag}
\tilde{a}_m\tilde{a}_k\tilde{a}_l
+
\tilde{a}_p^{\dag}\tilde{a}_q^{\dag}\tilde{a}_k\tilde{a}_l\delta_{mj}).
\end{eqnarray}
Combining all the terms that arise out of this procedure, and making use 
of the symmetries of the indices, equation~(\ref{eqn:n}) becomes
\begin{eqnarray}
\frac{d \hat{n}_p}{dt} &=&
\frac{2 \pi}{\hbar^2} \sum_{\stackrel{qmn}{ijkl}}
\langle pq|V|mn\rangle
\langle ij|V|kl\rangle
\delta(\omega_i + \omega_j - \omega_k - \omega_l)
\nonumber\\
&&
\times
\left[
\tilde{a}_i^{\dag}\tilde{a}_j^{\dag}\tilde{a}_p^{\dag}
\tilde{a}_l\tilde{a}_m\tilde{a}_n \delta_{kq}
+
\tilde{a}_i^{\dag}\tilde{a}_j^{\dag}\tilde{a}_q^{\dag}
\tilde{a}_l\tilde{a}_m\tilde{a}_n \delta_{kp}
+
\tilde{a}_i^{\dag}\tilde{a}_j^{\dag}
\tilde{a}_m\tilde{a}_n \delta_{kq}\delta_{lp}\right. \nonumber\\
&&
\left. - 
\tilde{a}_p^{\dag}\tilde{a}_q^{\dag}\tilde{a}_i^{\dag}
\tilde{a}_m\tilde{a}_k\tilde{a}_l \delta_{jn}
-
\tilde{a}_p^{\dag}\tilde{a}_q^{\dag}\tilde{a}_i^{\dag}
\tilde{a}_n\tilde{a}_k\tilde{a}_l \delta_{jm}
-
\tilde{a}_p^{\dag}\tilde{a}_q^{\dag}
\tilde{a}_k\tilde{a}_l \delta_{jn}\delta_{im}
\right].
\label{eqn:n2}
\end{eqnarray}

The next step is to take the expectation value of equation~(\ref{eqn:n2}), as we are
concerned with the time evolution of the {\em average} occupation of any individual
level.  In fact, it is most useful to take the ensemble average, which can be
written as
\begin{equation}
\langle \hat{A} \rangle = \mbox{Tr}(\hat{\rho} \hat{A}),
\label{eqn:thermal_average}
\end{equation}
where $\hat{\rho}$ is the density matrix of the system, and $\mbox{Tr}$ denotes
the trace operation.
We are left to calculate  quantities such as
$\langle
\tilde{a}_i^{\dag}\tilde{a}_j^{\dag}
\tilde{a}_m\tilde{a}_n 
\rangle
$.
To do so, we assume that our system is sufficiently 
near thermal equilibrium that we can
use  Wick's theorem \cite{Blaizot}.  This states that for any system with a
Hamiltonian that  is a quadratic form in creation and annihilation operators,
the ensemble average of any product of these is simply the contraction of
all possible pairings of the operators.  For example we have
\begin{eqnarray}
\langle
\tilde{a}_i^{\dag}\tilde{a}_j^{\dag}
\tilde{a}_m\tilde{a}_n 
\rangle
&=&
\langle \tilde{a}_i^{\dag}\tilde{a}_j^{\dag}\rangle
\langle\tilde{a}_m\tilde{a}_n \rangle +
 \langle\tilde{a}_i^{\dag} \tilde{a}_n \rangle 
\langle\tilde{a}_j^{\dag} \tilde{a}_m \rangle +
\langle\tilde{a}_i^{\dag} \tilde{a}_m \rangle 
\langle\tilde{a}_j^{\dag} \tilde{a}_n \rangle.
\end{eqnarray}
This relation is exact at thermal equilibrium, and should be a very good
approximation nearby.

We now make use of a further approximation, known as the  
random phase approximation (RPA), which states
that if the density matrix for the system is diagonal then
\begin{equation}
\langle \tilde{a}_i^{\dag}\tilde{a}_j^{\dag} \rangle = 
\langle \tilde{a}_i\tilde{a}_j \rangle = 0, \qquad
\langle \tilde{a}_i^{\dag}\tilde{a}_j \rangle = n_i \delta_{ij}.
\end{equation}
This will be an excellent approximation when there is no condensate present, as
the interaction Hamiltonian $\Hi$ is only a small
perturbation to the ideal gas Hamiltonian $H_0$.  Thus we have
\begin{eqnarray}
\langle
\tilde{a}_i^{\dag}\tilde{a}_j^{\dag}
\tilde{a}_m\tilde{a}_n 
\rangle
&=& n_m n_n (\delta_{in}\delta_{jm} + \delta_{im}\delta_{jn}),
\\
\langle\tilde{a}_i^{\dag}\tilde{a}_j^{\dag}\tilde{a}_p^{\dag}
\tilde{a}_l\tilde{a}_m\tilde{a}_n \rangle
& = &
n_l n_m n_n (
\delta_{il}\delta_{jm} \delta_{pn} + 
\delta_{il}\delta_{jn} \delta_{pm} + 
\delta_{im}\delta_{jl} \delta_{pn} \nonumber\\ 
&&\quad + \;\delta_{im}\delta_{jn} \delta_{pl} + 
\delta_{in}\delta_{jl} \delta_{pm} + 
\delta_{in}\delta_{jm} \delta_{pl}).
\end{eqnarray}
On substituting these relations into equation~(\ref{eqn:n2}) the final result is
\begin{mathletters}
\begin{eqnarray}
\frac{d n_p}{dt} &=&
\frac{4 \pi}{\hbar^2} \sum_{qmn}
|\langle pq|V|mn\rangle|^2
\delta(\omega_p + \omega_q - \omega_m - \omega_n)\nonumber \\
&&\;\;\;\;\;\;\;\;\;\times  
\left[(n_p + 1)(n_q + 1) n_m n_n - n_p n_q (n_m+1)(n_n+1)\right]
\label{eqn:QBEa}
\\
&+&
\frac{8 \pi}{\hbar^2} \sum_{qmn}
\langle pq|V|pn\rangle\langle mn|V|mq\rangle
n_p n_m (n_n - n_q) \delta(\omega_n - \omega_q)
\label{eqn:qbe1}
\\
&+&
\frac{8 \pi}{\hbar^2} \sum_{qmn}
\langle pq|V|qn\rangle\langle mn|V|mp\rangle
n_q n_m (n_n - n_p) \delta(\omega_n - \omega_p).
\label{eqn:qbe2}
\end{eqnarray}
\end{mathletters}
The first part of this expression (\ref{eqn:QBEa}) is the standard quantum
Boltzmann equation; however, the lines of (\ref{eqn:qbe1}) and (\ref{eqn:qbe2}) do not
appear in most definitions of the QBE.
We would like to note the following about these terms:
\begin{enumerate}
\item{The scattering processes described by the matrix elements in these terms
involves a third particle,
and hence these collision terms are of higher order than  those described by
(\ref{eqn:QBEa}).}
\item{If we calculate the matrix elements using the contact potential
approximation in the homogeneous limit, then these terms become
\begin{mathletters}
\begin{eqnarray}
\hspace*{-10mm}\mbox{(\protect\ref{eqn:qbe1})}&\rightarrow&
\frac{8 \pi U_0^2}{\hbar^2 \Omega^2} \sum_{qmn}
\delta({\bf k}_q - {\bf k}_n)^2
n_p n_m (n_n - n_q) \delta(\omega_n - \omega_q),
\label{eqn:hqbe1}
\\
\hspace*{-10mm}\mbox{(\protect\ref{eqn:qbe2})}&\rightarrow&
\frac{8 \pi U_0^2}{\hbar^2 \Omega^2}\sum_{qmn}
\delta({\bf k}_p - {\bf k}_n)^2
n_q n_m (n_n - n_p) \delta(\omega_n - \omega_p).
\label{eqn:hqbe2}
\end{eqnarray}
\end{mathletters}
The delta functions in momentum are equivalent to Kronecker delta functions in the
quantum labels for the system, and hence these terms vanish.
}
\item{For an ergodic system where the occupation of a level
depends only on its energy,  these terms are again identically zero.}
\item{The delta functions in frequency depend
on only two of the particle indices, rather than four as for 
equation~(\ref{eqn:QBEa}).  This means there will be far fewer matches for 
(\ref{eqn:qbe1}) and (\ref{eqn:qbe2}), and therefore these can be considered
surface terms that become small in the thermodynamic limit.}
\end{enumerate}
We are therefore justified in neglecting
these terms, and are left with the usual quantum
Boltzmann equation
\begin{eqnarray}
\frac{d n_p}{dt} &=&
\frac{4 \pi}{\hbar} \sum_{qmn}
|\langle pq|V|mn\rangle|^2
\delta(\epsilon_p + \epsilon_q - \epsilon_m - \epsilon_n)\nonumber \\
&&\;\;\;\;\;\;\;\;\;\times  
\bigg\{(n_p + 1)(n_q + 1) n_m n_n - n_p n_q (n_m+1)(n_n+1)\bigg\}.
\label{eqn:QBE}
\end{eqnarray}
\subsubsection*{Validity}
We summarize the validity conditions for the QBE as:
\begin{enumerate}
\item{The Markov approximation must be valid such that correlations induced by
collisions are unimportant.}
\item{The system should close enough to equilibrium such that the factorization of
Wick's theorem is valid.}
\item{There must be a good basis such that the RPA is valid.  In our derivation we
have assumed that $\Hi$ should be a perturbation to the system for this to hold.
However, if the average effect of $\Hi$ can be absorbed into $\Ho$ to form an
effective Hamiltonian with a good basis,
then the QBE derivation may still be valid.  It must be noted, however, 
 that the Markov 
approximation may not be valid in this regime as the mean collision time will
be much reduced.}
\end{enumerate}

\section{The GPE kinetic equation}\label{GPEkinetic}
It is interesting to consider the kinetic equation that would result if we
assume that the GPE is a good description of the system of interest.
We expand the time dependent wave function as
\begin{equation}
\psi({\bf x},t) = \sum_{k} \tilde{c}_k \phi_k({\bf x}) e^{-i \omega_k t},
\label{eqn:expsi}
\end{equation}
where the $\{\tilde{c}_k\}$ are slowly varying. 
Substituting equation~(\ref{eqn:expsi}) into the time-dependent GPE
(\ref{eqn:gpe})
and performing the operation
$\int d^3{\bf x} \phi_p^*({\bf x})$ on both sides results in the basis set
representation of the GPE
\begin{equation}
i \hbar \frac{d \tilde{c}_p}{dt}
=
\sum_{qmn}\langle pq|V|mn\rangle
\tilde{c}_q^*\tilde{c}_m\tilde{c}_n
e^{i(\omega_p + \omega_q - \omega_m - \omega_n)t}.
\label{eqn:ct}
\end{equation}
We note that equation~(\ref{eqn:ct}) is identical in form to the basis set
 equation of motion for the
Bose field (\ref{eqn:at}) but for the replacement
$\tilde{a}_k\leftrightarrow\tilde{c}_k$. 
In fact, equation~(\ref{eqn:ct}) could be derived directly from
equation~(\ref{eqn:at}) by expanding the annihilation operators as $\hat{a}_k = c_k
+ \hat{\delta}_k $ and then neglecting the quantum fluctuation
terms $\hat{\delta}_k $.

We can now carry out the same procedure on (\ref{eqn:ct})
as was applied to equation~(\ref{eqn:at})
in the derivation of  the QBE. 
The only difference is  that we are now manipulating 
{\em c}-numbers rather than operators, and so 
any terms arising from commutators in the previous treatment will
not appear.  This means that the terms of the form $ \tilde{a}^{\dag}\tilde{a}^{\dag}
\tilde{a}\tilde{a}$ will disappear from equation~(\ref{eqn:n2}), leaving only terms
involving six {\em c}-numbers.  Writing $n_p$ for $\tilde{c}_p^*\tilde{c}_p$,
the resulting GPE kinetic equation is
\begin{eqnarray}
\frac{d n_p}{dt} &=&
\frac{4 \pi U_0^2}{\hbar} \sum_{qmn}
|\langle pq|mn\rangle|^2
\delta(\epsilon_p + \epsilon_q - \epsilon_m - \epsilon_n)\nonumber \\
&&\;\;\;\;\;\;\;\;\;\times  
\bigg\{(n_p + n_q ) n_m n_n - n_p n_q (n_m + n_n)\bigg\}
\label{eqn:GPE_ke}
\end{eqnarray}
which is exactly the same form as the QBE (\ref{eqn:QBE}) except that the
spontaneous collision terms are excluded.  This equation was first considered by
Svistunov in a study of the formation of a condensate in a 
weakly-interacting Bose gas \cite{boris}.

Some of the approximations made in the derivation of the GPE kinetic equation
may not hold in the presence of a condensate.  In particular, 
the assumption of
correlations being unimportant on the scale of the 
 collision time is unlikely
to be valid.  The GPE kinetic equation does, however, give us an understanding
of the collision processes that are described by the full GPE.

From equation~(\ref{eqn:GPE_ke}) we can see that the GPE contains stimulated
collision processes only.  To understand this, consider the collision $p + q \rightarrow m
+ n$.  This process will only be represented by the GPE
if one of the levels $(m,n)$ is already occupied.  This is in contrast to
the QBE, for which the term in the curly brackets of equation~(\ref{eqn:QBE}) can
be written
\begin{equation}
\bigg\{(1 + n_p + n_q ) n_m n_n - n_p n_q (1 + n_m + n_n)\bigg\}.
\end{equation}
Thus, due to the neglect of the quantum nature of the modes, the GPE can only accurately describe the evolution  and interaction of modes which satisfy 
$n_p \gg 1$, such that  $(1 + n_p + n_q ) \approx (n_p + n_q ) $.

\section{Incoherent region equations}\label{a:incoherent} 

In this appendix we write out the Heisenberg equations of motion for all the
operator combinations that appear in the basis set version of the FTGPE 
(\ref{eqn:FTGPEbasis}). The single operator term is

\begin{mathletters}
\begin{eqnarray}
i \hbar \,\frac{d \hat{a}_{p}}{d t} &=& 
\hbar \omega_p \hat{a}_p
\label{aa}
 \\
&+&
 \langle p \psi |V|\psi \psi \rangle
\label{ab}
\\
&+&
\sum_{q}\langle p q |V|\psi \psi \rangle \hat{a}_q^{\dag}
\label{ac}
\\
&+&
2 \sum_{ m }\langle p \psi |V| m \psi \rangle \hat{a}_m
\label{ad}
\\
&+&
\sum_{ mn}\langle p \psi |V|m n \rangle \hat{a}_m \hat{a}_n
\label{ae}
\\
&+&
2 \sum_{qm}\langle p q |V|m \psi \rangle
\hat{a}_q^{\dag}\hat{a}_m
\label{af}
\\
&+&
\sum_{qmn}\langle p q |V|m n \rangle \hat{a}_q^{\dag}\hat{a}_m\hat{a}_n.
\label{ag}
\end{eqnarray}
\label{eqn:a2_appendix}
\end{mathletters}
 The equation of
motion for $\hat{a}_p^{\dag}$ is simply the hermitian conjugate of 
(\ref{eqn:a2_appendix}).
The other equations of motion can be found either by calculating the commutator
with the Hamiltonian, or using the chain rule. 
We have
\begin{mathletters}
\begin{eqnarray}
i \hbar \frac{d (\hat{a}_m \hat{a}_n)}{d t} 
&=& 
\hbar (\omega_m + \omega_n) \hat{a}_m \hat{a}_n\\
&+& 
\langle m \psi |V|\psi \psi \rangle \hat{a}_n +
 \langle n \psi |V|\psi \psi \rangle \hat{a}_m\\
&+&  
\langle m n  |V|\psi \psi \rangle\\
&+&
\sum_k \left[ \langle kn  |V|\psi \psi \rangle \hat{a}_k^{\dag}\hat{a}_m +
\langle km  |V|\psi \psi \rangle \hat{a}_k^{\dag}\hat{a}_n \right]\\
&+& 
2 \sum_k \left[\langle n \psi  |V|k \psi \rangle \hat{a}_k\hat{a}_m +
\langle m \psi  |V|k \psi \rangle \hat{a}_k\hat{a}_n \right]\\
&+&
\sum_{kj} \left[\langle m \psi |V|kj \rangle \hat{a}_n\hat{a}_k\hat{a}_j
+ \langle n \psi |V|kj \rangle \hat{a}_m \hat{a}_k \hat{a}_j\right]\\
&+&
2 \sum_{k}\langle m n |V|k\psi \rangle\hat{a}_k\\
&+&
2\sum_{kj} \left[\langle km |V|j \psi \rangle \hat{a}^{\dag}_k \hat{a}_j \hat{a}_n
+ \langle kn |V|j \psi \rangle \hat{a}^{\dag}_k \hat{a}_j \hat{a}_m\right]\\
&+&
\sum_{kj} \langle mn |V|kj \rangle \hat{a}_k \hat{a}_j\\
&+&
\sum_{qkj} \left[\langle mq |V|kj \rangle \hat{a}^{\dag}_q\hat{a}_n \hat{a}_k\hat{a}_j
+\langle nq |V|kj \rangle \hat{a}^{\dag}_q\hat{a}_m\hat{a}_k \hat{a}_j\right],
\end{eqnarray}
\label{eqn:aa}
\end{mathletters}
\begin{mathletters}
\begin{eqnarray}
i \hbar \frac{d (\hat{a}^{\dag}_m \hat{a}_n)}{d t} 
&=& \hbar (\omega_n - \omega_m) \hat{a}^{\dag}_m \hat{a}_n\\
&+&
\langle n\psi  |V|\psi \psi \rangle \hat{a}^{\dag}_m -
\langle \psi \psi  |V| m \psi  \rangle \hat{a}_n\\
&+&
\sum_{k} \left[\langle k n |V|\psi\psi \rangle \hat{a}^{\dag}_k \hat{a}_m^{\dag} 
- \langle \psi \psi |V| k m \rangle \hat{a}_k \hat{a}_n\right]\\
&+&
2 \sum_{k} \left[\langle  n \psi |V|k \psi \rangle \hat{a}^{\dag}_m \hat{a}_k 
- \langle k \psi |V|m \psi \rangle \hat{a}_k^{\dag} \hat{a}_n\right]\\
&+&
\sum_{kj} \left[\langle  n \psi |V|k j \rangle \hat{a}^{\dag}_m \hat{a}_k \hat{a}_j
- \langle k j |V|m \psi \rangle \hat{a}_k^{\dag} {a}_j^{\dag}\hat{a}_n\right]\\
&+&
2 \sum_{kj} \left[\langle k n  |V| j \psi \rangle \hat{a}^{\dag}_m \hat{a}_k^{\dag} \hat{a}_j
- \langle j \psi |V|km \rangle \hat{a}_j^{\dag} {a}_k\hat{a}_n\right]\\
&+&
\sum_{qkj} \left[\langle q n  |V| kj \rangle \hat{a}^{\dag}_m \hat{a}_q^{\dag} \hat{a}_k \hat{a}_j
- \langle kj |V| qm \rangle \hat{a}_j^{\dag} {a}_k^{\dag} \hat{a}_q \hat{a}_n
\right],
\end{eqnarray}
\label{eqn:adaga_appendix}
\end{mathletters}
and
\begin{mathletters}
\begin{eqnarray}
i \hbar \frac{d (\hat{a}^{\dag}_q  \hat{a}_m \hat{a}_n)}{d t} 
&=& i \hbar \left[ \hat{a}^{\dag}_q \frac{d (\hat{a}_m \hat{a}_n)}{d t} + 
\frac{d (\hat{a}^{\dag}_q )}{d t}\hat{a}^{\dag}_m \hat{a}_n \right]
\nonumber\\
&=& \hbar (\omega_m + \omega_n - \omega_q) \hat{a}^{\dag}_q \hat{a}_m \hat{a}_n\\
&+& 
\langle m \psi |V|\psi \psi \rangle \hat{a}_q^{\dag} \hat{a}_n +
\langle n \psi |V|\psi \psi \rangle \hat{a}_q^{\dag} \hat{a}_m\\
&+&  
\langle m n  |V|\psi \psi \rangle \hat{a}_q^{\dag} -
\langle \psi \psi |V|q \psi \rangle \hat{a}_m \hat{a}_n\\
&+&
\sum_k \left[ \langle kn  |V|\psi \psi \rangle \hat{a}_q^{\dag} \hat{a}_k^{\dag}\hat{a}_m +
\langle km  |V|\psi \psi \rangle \hat{a}_q^{\dag} \hat{a}_k^{\dag}\hat{a}_n \right. \\
&& \left.-\langle \psi \psi |V| kq \rangle \hat{a}_k \hat{a}_m \hat{a}_n \right]\\
&+& 
2 \sum_k \left[ \langle n \psi  |V|k \psi \rangle \hat{a}_q^{\dag} \hat{a}_k\hat{a}_m +
\langle m \psi  |V|k \psi \rangle \hat{a}_q^{\dag} \hat{a}_k\hat{a}_n \right.\\
&& \left. -
\langle k \psi  |V|q \psi \rangle \hat{a}_k^{\dag} \hat{a}_m\hat{a}_n  \right]\\
&+&
\sum_{kj} \left[\langle m \psi |V|kj \rangle \hat{a}_q^{\dag} \hat{a}_n\hat{a}_k\hat{a}_j
+ \langle n \psi |V|kj \rangle \hat{a}_q^{\dag} \hat{a}_m \hat{a}_k \hat{a}_j \right. \\
&& \left.- \langle kj |V|q \psi \rangle \hat{a}_k^{\dag} {a}_j^{\dag} \hat{a}_m \hat{a}_n \right]\\
&+&
2\sum_{k}\langle m n |V|k\psi \rangle\hat{a}_q^{\dag} \hat{a}_k\\
&+&
2\sum_{kj} \left[ \langle km |V|j \psi \rangle \hat{a}_q^{\dag}\hat{a}^{\dag}_k \hat{a}_j \hat{a}_n
+\langle kn |V|j \psi \rangle \hat{a}_q^{\dag}\hat{a}^{\dag}_k \hat{a}_j \hat{a}_m \right.\\
&& \left. - \langle j \psi  |V| k q \rangle \hat{a}_k^{\dag}\hat{a}_j \hat{a}_m \hat{a}_n \right]\\
&+&
\sum_{kj} \langle mn |V|kj \rangle \hat{a}_q^{\dag} \hat{a}_k \hat{a}_j\\
&+&
\sum_{rkj} \left[ \langle mr |V|kj \rangle \hat{a}^{\dag}_q\hat{a}^{\dag}_r\hat{a}_n \hat{a}_k\hat{a}_j
+ \langle nr |V|kj \rangle \hat{a}^{\dag}_q\hat{a}^{\dag}_r\hat{a}_m\hat{a}_k \hat{a}_j\right.\\
&&\left.- \langle kj |V| rq \rangle
\hat{a}^{\dag}_k\hat{a}^{\dag}_j\hat{a}_r\hat{a}_m \hat{a}_n \right].
\end{eqnarray}
\label{eqn:adagaa_appendix}
\end{mathletters}

\section{Incoherent region kinetic equation}\label{incoherentKE}

In this section we present the kinetic equation of motion that can be derived
for the incoherent region, using the same methods as in the derivation of the
QBE in appendix~\ref{sec:qbe} and the techniques outlined in section~\ref{ftgpe_terms}.  We find
\begin{mathletters}
\begin{eqnarray}
\frac{d n_p }{dt} 
&=&
\frac{2\pi}{\hbar}\sum_{q \p \q} |\langle pq  |V|\p \q \rangle|^2 
\delta(\hbar\omega_{pq \p \q})|c_{\p} c_{\q}|^2 (n_p + n_q + 1)
\label{kanom}
\\
&+&
\frac{8\pi}{\hbar}
\sum_{\p \q m}|\langle p \p |V| m \q \rangle|^2\delta(\hbar\omega_{p \p m \q})
|c_{\p} c_{\q}|^2 (n_m - n_p)
\label{kscatt}
\\
&+&
\frac{4\pi}{\hbar}\sum_{\p mn}
|\langle p \p |V|m n \rangle|^2\delta(\hbar\omega_{p \p m n})
\nonumber\\
&&\qquad\times|c_{\p} |^2\left\{n_m n_n - n_p( 1 +  n_m + n_n)\right\}
\label{kgrow1}
\\
&+&
\frac{8\pi}{\hbar}
\sum_{\p qm}|\langle p q |V|m \p \rangle|^2
\delta(\hbar\omega_{pqm\p})\nonumber
\\
&&\qquad \times \; |c_{\p} |^2\left\{ (n_p + n_q + 1)n_m - n_p n_q
\right\}
\label{kgrow2}
\\
&+&
\frac{4\pi}{\hbar}
\sum_{qmn}|\langle p q |V|m n \rangle |^2
\delta(\hbar\omega_{pqmn}) \nonumber\\
&&\qquad \times\;\left\{(1 + n_p)(1 + n_q)n_m n_n  - n_p n_q (1 + n_m)(1 + n_n)
\right\}.
\label{kqbe}
\end{eqnarray}
\end{mathletters}
We can recognize each of these terms from our previous discussions:
\begin{enumerate}
\item{
The term (\ref{kanom}) is from the anomalous term, and is only non-zero
when we consider the collision of multiple condensates.
}
\item{
The line (\ref{kscatt}) describes the scattering processes.
}
\item{
The two terms (\ref{kgrow1}) and~(\ref{kgrow2})
are due to the forward and backward growth terms of the coherent region.}

\item{
The line  (\ref{kqbe}) is simply the QBE for the incoherent region.}
\end{enumerate}

It may seem surprising that there is no contribution from the
linear terms discussed in the previous section.  This is because the rates for
each of the forward and backward processes contain only stimulated terms and so
they cancel.
The kinetic equation for the incoherent region is thus the
usual QBE but with additional couplings to the coherent region
whose physical meanings can be understood from
the corresponding terms in section~\ref{ftgpe_terms}.

\section{Conservation of normalization in the PGPE}\label{PGPEnorm}

The PGPE conserves normalization and energy because the effective projected
Hamiltonian is hermitian. The nonlinear term of the GPE can be considered to
describe interactions between two particles, and as such  there can be
collisions in which two coherent atoms collide and one is ejected from the
coherent region.  However, the projector excludes these terms from the equation
of motion which we now demonstrate.

Consider the equation of motion 
in a basis set.  By substituting
$\psi({\bf x})  = \sum_{{ k} \in C} c_{ k} \phi_{ k}({\bf x})$
 into equation~(\ref{eqn:pgpe}) and performing the operation 
$\int d^3 {\bf x} \phi_p({\bf x})$ on both sides we find
\begin{equation}
i \hbar \frac{d c_p}{d t}
= \hbar \omega_{p} c_p + U_0
\hat{\mathcal{P}}
\sum_{qmn \in C} c^*_q c_m c_n \langle pq |mn \rangle.
\label{eqn:basis_set}
\end{equation}
If the state $p \in C$ then  all four labels are from the coherent region and
there is no transfer of population outside the region.
For $p \notin C$ the  matrix element
$\langle pq |mn \rangle$ is not zero necessarily, and therefore it seems 
collisions between states
from within the coherent region can transfer population outside of $C$.
However, we should not be considering the equations of motion for amplitudes
$p \notin C$ in the first place, 
as they not in the definition of the  wave function  $\psi({\bf x})$.
The projection operation is therefore performed implicitly by
the basis set representation.

Numerically solving the GPE using a basis set method  requires a triple
summation over indices, which is a very time-consuming operation.  This
suggests that we should instead use the spatial representation of
equation~(\ref{eqn:pgpe}), where the nonlinear term is local.  However, any spatial
grid that is fine enough to provide a good representation of all  the states
within $C$  will also be able to represent modes outside the region $C$.  From 
equation~(\ref{eqn:basis_set}) we can see that this will cause  population to be
transferred outside of $C$, and so in this case we need to consider the
projection operation explicitly.  

Another method of approaching this issue
is to assume that {\em all} modes in the problem can
be represented by the GPE,
but artificially choose part of the system to be the
coherent region.  Thus the field operator can be written as 
\begin{equation}
\hat{\Psi}({\bf x}) \approx \psi({\bf x}) + \eta({\bf x}),
\end{equation}
where {\em both} fields are classical.
Substituting this into the interaction part of the Hamiltonian (\ref{eqn:Hi})
and using the
contact potential approximation, we have
\begin{eqnarray}
\Hi / U_0 &=&  \hat{H}_4 + \hat{H}_3+ \hat{H}_2+ \hat{H}_1+ \hat{H}_0,
\\
\hat{H}_4 & = &  \frac{1}{2}\psi^* \psi^* \psi \psi ,
\label{h4}\\
\hat{H_3} & = &  \psi^* \psi^* \psi \eta +  \psi^* \eta^* \psi \psi,  
\label{h3}\\
\hat{H_2} & = & \frac{1}{2}\psi^* \psi^* \eta \eta + 2 \psi^* \eta^* \psi \eta
+ \frac{1}{2}\eta^* \eta^* \psi \psi  ,
\label{h2}\\
\hat{H_1} & = &  \psi^* \eta^* \eta \eta +   \eta^* \eta^* \psi \eta  ,
\label{h1}\\
\hat{H_0} & = & \frac{1}{2}\eta^* \eta^* \eta \eta , 
\label{h0}
\label{eqn:split_H}
\end{eqnarray}
where we have dropped all space and time labels for clarity, and have
divided the Hamiltonian into five terms
 depending on the number of coherent region
fields they contain. 
Considering the Hamiltonian in this form we can easily
interpret each of the terms.  Each
$\psi^*$ creates a particle in the coherent region, and each $\psi$ removes a
particle from the coherent region.  The $\eta^*$ and $\eta$ perform the same
operation outside the region $C$.  This allows us to identify which
processes each of the terms in the Hamiltonian contribute
to the equations of motion for both $\psi$
and $\eta$.

We can now derive equations of motion for $\psi$ and $\eta$ by using functional
differentiation. We find
\begin{equation}
i \hbar \frac{\partial \psi}{\partial t} = 
\hat{\mathcal{P}}\frac{\partial \hat{H}}{\partial \psi^*},
\;\;\;\;\;
i \hbar\frac{\partial \eta}{\partial t} = \hat{\mathcal{Q}}\frac{\partial \hat{H}}{\partial \eta^*}.
\label{eqn:functionald}
\end{equation}
As an example, let us consider the contribution to these equations for all
interactions involving three coherent  and one incoherent state described by
$\hat{H}_3$.
We find from (\ref{h3}) and (\ref{eqn:functionald})
\begin{eqnarray}
i \hbar \frac{\partial \psi}{\partial t}
&\sim& \hat{\mathcal{P}} \left( 2 |\psi|^2 \eta +  \eta^* \psi^2 \right),\\
i \hbar\frac{\partial \eta}{\partial t}
&\sim &\hat{\mathcal{Q}} \left(  |\psi|^2 \psi\right).
\end{eqnarray}
The results of carrying out this operation for all particle processes are
summarized in table~\ref{tab:ftgpe}.  
The equations of motion for the system  will together conserve
both energy and normalization if  {\em all} terms in any row of the
table are included, as this correctly accounts for all forward and backward
processes of the same order.
We have confirmed this numerically by carrying out coupled
simulations of $\psi$ and $\eta$ and including only some of these terms.

We can now see that if we want an equation describing
interactions involving four coherent states, but neglecting all processes
involving incoherent
particles, then this is simply the PGPE.

\begin{table}\centering
\caption{Classical FTGPE divided into terms representing 
physical processes involving $n$ coherent states}
\begin{tabular}{ccc}
\begin{tabular}{c}
No. of\\
coherent\\
states\\
\end{tabular} &
$\displaystyle i \hbar \bigg(\frac{\partial \psi}{\partial t}\bigg) = 
\hat{\mathcal{P}} \times \ldots$&
$\displaystyle
i\hbar  \bigg(\frac{\partial \eta}{\partial t}\bigg) = \hat{\mathcal{Q}} \times\ldots $ \\
\hline
4&$|\psi|^2\psi$&$0$\\
3&$+2 |\psi|^2 \eta +  \eta^* \psi^2 $&$+|\psi|^2\psi$\\
2&$+2\psi |\eta|^2 + \psi^* \eta^2$&$+2 |\psi|^2 \eta +  \eta^* \psi^2$\\
1&$+|\eta|^2 \eta$&$+2\psi |\eta|^2 + \psi^* \eta^2$\\
0&$+0$&$+|\eta|^2 \eta$\\
\end{tabular}
\label{tab:ftgpe}
\end{table}

\end{appendix}

\end{document}